%% file: main.tex
\documentclass[12pt,a4paper]{article}
\usepackage{ifthen} % for conditional statements
\newboolean{pdflatex}
\setboolean{pdflatex}{true} % False for eps figures 

\newboolean{articletitles}
\setboolean{articletitles}{true} % False removes titles in references

\newboolean{uprightparticles}
\setboolean{uprightparticles}{false} %True for upright particle symbols

\newboolean{inbibliography}
\setboolean{inbibliography}{false} %True once you enter the bibliography

\newboolean{prlversion}
\setboolean{prlversion}{false} 

\newboolean{wordcount}
\setboolean{wordcount}{false} % False for normal usage; true for wordcount.sh

\def\paperauthors{LHCb collaboration} % Leave as is for PAPER and CONF

\def\paperasciititle{} % Set ASCII title here

\def\papertitle{Measurement of $f_s / f_u$ variation with proton-proton collision energy and $B$-meson kinematics}

\def\paperkeywords{{High Energy Physics}, {LHCb}} % Comma separated list

\def\papercopyright{\the\year\ CERN for the benefit of the LHCb collaboration} % new since 9/Apr/2018
\def\paperlicence{CC-BY-4.0 licence}
\def\paperlicenceurl{https://creativecommons.org/licenses/by/4.0/}

\input{preamble}
\usepackage{longtable} % only for template; not usually to be used in PAPERs
\usepackage{booktabs}
\begin{document}

%%%%%%%%%%%%%%%%%%%%%%%%%
%%%%% Title     %%%%%%%%%
%%%%%%%%%%%%%%%%%%%%%%%%%
\renewcommand{\thefootnote}{\fnsymbol{footnote}}
\setcounter{footnote}{1}

%\onecolumn
\input{title-LHCb-PAPER}

%\twocolumn
% %%%%%%%%%%%%% ---------

\renewcommand{\thefootnote}{\arabic{footnote}}
\setcounter{footnote}{0}

%%%%%%%%%%%%%%%%%%%%%%%%%%%%%%%%
%%%%%  Table of Content   %%%%%%
%%%%%%%%%%%%%%%%%%%%%%%%%%%%%%%%
%%%% Uncomment next 2 lines if desired
%\tableofcontents
%\cleardoublepage

%%%%%%%%%%%%%%%%%%%%%%%%%
%%%%% Main text %%%%%%%%%
%%%%%%%%%%%%%%%%%%%%%%%%%

\pagestyle{plain} % restore page numbers for the main text
\setcounter{page}{1}
\pagenumbering{arabic}

%% Uncomment during review phase. 
%% Comment before a final submission.
%\linenumbers

\input{introduction}

\input{detector}

\input{selection}

\input{massfit}

\input{efficiencies}

\input{syst}

\input{results}

% Comment this in for paper drafts; do not include this in analysis note and conference reports
\input{acknowledgements}

\addcontentsline{toc}{section}{References}
\setboolean{inbibliography}{true}
\bibliographystyle{LHCb}
\bibliography{main,standard,LHCb-PAPER,LHCb-CONF,LHCb-DP,LHCb-TDR}

\appendix 
\input{supplementary-app}

\clearpage

\newpage
\input{LHCb_Authorship_01-Oct-2019.tex}

\end{document}

%% file: preamble.tex
\usepackage[top=1in, bottom=1.25in, left=1in, right=1in]{geometry}

\usepackage{microtype}
\usepackage{lineno}  % for line numbering during review
\usepackage{xspace} % To avoid problems with missing or double spaces after
\usepackage{caption} %these three command get the figure and table captions automatically small

\usepackage{graphicx}  % to include figures (can also use other packages)
\usepackage{subcaption}  %subfigures

\usepackage{color}
\usepackage{colortbl}
\graphicspath{{./figs/}} % Make Latex search fig subdir for figures
\DeclareGraphicsExtensions{.pdf,.PDF,png,.PNG}

\usepackage{amsmath} % Adds a large collection of math symbols
\usepackage{amssymb}
\usepackage{amsfonts}
\usepackage{upgreek} % Adds in support for greek letters in roman typeset
\usepackage{overpic}
\newcommand*\patchAmsMathEnvironmentForLineno[1]{%
\expandafter\let\csname old#1\expandafter\endcsname\csname #1\endcsname
\expandafter\let\csname oldend#1\expandafter\endcsname\csname
end#1\endcsname
 \renewenvironment{#1}%
   {\linenomath\csname old#1\endcsname}%
   {\csname oldend#1\endcsname\endlinenomath}%
}
\newcommand*\patchBothAmsMathEnvironmentsForLineno[1]{%
  \patchAmsMathEnvironmentForLineno{#1}%
  \patchAmsMathEnvironmentForLineno{#1*}%
}
\AtBeginDocument{%
\patchBothAmsMathEnvironmentsForLineno{equation}%
\patchBothAmsMathEnvironmentsForLineno{align}%
\patchBothAmsMathEnvironmentsForLineno{flalign}%
\patchBothAmsMathEnvironmentsForLineno{alignat}%
\patchBothAmsMathEnvironmentsForLineno{gather}%
\patchBothAmsMathEnvironmentsForLineno{multline}%
\patchBothAmsMathEnvironmentsForLineno{eqnarray}%
}

\usepackage{hyperxmp}

\usepackage[pdftex,
            pdfauthor={\paperauthors},
            pdftitle={\paperasciititle},
            pdfkeywords={\paperkeywords},
            pdfcopyright={Copyright (C) \papercopyright},
            pdflicenseurl={\paperlicenceurl}]{hyperref}

\usepackage[colorinlistoftodos,textsize=scriptsize]{todonotes}

\usepackage[all]{hypcap} % Internal hyperlinks to floats.

\input{lhcb-symbols-def} % Add in the predefined LHCb symbols
\input{our-symbols-def}

\input{thesisstyle}

\usepackage{cite} % Allows for ranges in citations
\usepackage{mciteplus}

%% file: lhcb-symbols-def.tex
\usepackage{xspace} 
\usepackage{upgreek}

\def\MagUp {\mbox{\em Mag\kern -0.05em Up}\xspace}

\ifthenelse{\boolean{uprightparticles}}%
{

 \def\Pmu         {\ensuremath{\upmu}\xspace}

 \def\Ppsi        {\ensuremath{\uppsi}\xspace}

 \def\PDelta      {\ensuremath{\Delta}\xspace}                 
 \def\PXi         {\ensuremath{\Xi}\xspace}                 
 \def\PLambda     {\ensuremath{\Lambda}\xspace}                 
 \def\PSigma      {\ensuremath{\Sigma}\xspace}                 
 \def\POmega      {\ensuremath{\Omega}\xspace}                 
 \def\PUpsilon    {\ensuremath{\Upsilon}\xspace}

 \def\PB      {\ensuremath{\mathrm{B}}\xspace}                 
                  
 \def\PD      {\ensuremath{\mathrm{D}}\xspace}

 \def\PJ      {\ensuremath{\mathrm{J}}\xspace}                 
 \def\PK      {\ensuremath{\mathrm{K}}\xspace}

 \def\PZ      {\ensuremath{\mathrm{Z}}\xspace}                 
                  
 \def\Pb      {\ensuremath{\mathrm{b}}\xspace}

 \def\Pi      {\ensuremath{\mathrm{i}}\xspace}

 \def\Pp      {\ensuremath{\mathrm{p}}\xspace}

 \def\Ps      {\ensuremath{\mathrm{s}}\xspace}

 \def\thebaroffset{0.0em}
}
{

 \def\Pmu         {\ensuremath{\mu}\xspace}

 \def\Ppsi        {\ensuremath{\psi}\xspace}                 
                  
 \mathchardef\PDelta="7101
 \mathchardef\PXi="7104
 \mathchardef\PLambda="7103
 \mathchardef\PSigma="7106
 \mathchardef\POmega="710A
 \mathchardef\PUpsilon="7107
                  
 \def\PB      {\ensuremath{B}\xspace}                 
                  
 \def\PD      {\ensuremath{D}\xspace}

 \def\PJ      {\ensuremath{J}\xspace}                 
 \def\PK      {\ensuremath{K}\xspace}

 \def\PZ      {\ensuremath{Z}\xspace}                 
                  
 \def\Pb      {\ensuremath{b}\xspace}

 \def\Pi      {\ensuremath{i}\xspace}

 \def\Pp      {\ensuremath{p}\xspace}

 \def\Ps      {\ensuremath{s}\xspace}

 \def\thebaroffset{0.18em}
}
\newcommand{\offsetoverline}[2][\thebaroffset]{\kern #1\overline{\kern -#1 #2}}%

\makeatletter
\ifcase \@ptsize \relax% 10pt
  \newcommand{\miniscule}{\@setfontsize\miniscule{4}{5}}% \tiny: 5/6
\or% 11pt
  \newcommand{\miniscule}{\@setfontsize\miniscule{5}{6}}% \tiny: 6/7
\or% 12pt
  \newcommand{\miniscule}{\@setfontsize\miniscule{5}{6}}% \tiny: 6/7
\fi
\makeatother

\DeclareRobustCommand{\optbar}[1]{\shortstack{{\miniscule (\rule[.5ex]{1.25em}{.18mm})}
  \\ [-.7ex] $#1$}}

\def\mup        {{\ensuremath{\Pmu^+}}\xspace}
\def\mun        {{\ensuremath{\Pmu^-}}\xspace} % muon negative (\mum is taken)

\def\Z      {{\ensuremath{\PZ}}\xspace}

\def\squark    {{\ensuremath{\Ps}}\xspace}

\def\bquark    {{\ensuremath{\Pb}}\xspace}
\def\bquarkbar {{\ensuremath{\overline \bquark}}\xspace}

\def\kaon    {{\ensuremath{\PK}}\xspace}
\def\Kbar    {{\ensuremath{\offsetoverline{\PK}}}\xspace}

\def\KorKbar {\kern \thebaroffset\optbar{\kern -\thebaroffset \PK}{}\xspace}

\def\Kp      {{\ensuremath{\kaon^+}}\xspace}

\def\KS      {{\ensuremath{\kaon^0_{\mathrm{S}}}}\xspace}
\def\KL      {{\ensuremath{\kaon^0_{\mathrm{L}}}}\xspace}

\def\Kstarzb {{\ensuremath{\Kbar{}^{*0}}}\xspace}

\def\DorDbar {\kern \thebaroffset\optbar{\kern -\thebaroffset \PD}\xspace}

\def\B       {{\ensuremath{\PB}}\xspace}

\def\BorBbar {\kern \thebaroffset\optbar{\kern -\thebaroffset \PB}\xspace}

\def\Bd      {{\ensuremath{\B^0}}\xspace}

\def\BdorBdbar {\kern \thebaroffset\optbar{\kern -\thebaroffset \Bd}\xspace}
\def\Bu      {{\ensuremath{\B^+}}\xspace}

\def\Bs      {{\ensuremath{\B^0_\squark}}\xspace}

\def\BsorBsbar {\kern \thebaroffset\optbar{\kern -\thebaroffset \Bs}\xspace}

\def\jpsi     {{\ensuremath{{\PJ\mskip -3mu/\mskip -2mu\Ppsi\mskip 2mu}}}\xspace}

\def\Y#1S{\ensuremath{\PUpsilon{(#1S)}}\xspace}

\def\proton      {{\ensuremath{\Pp}}\xspace}
\def\antiproton  {{\ensuremath{\overline \proton}}\xspace}

\def\Lz          {{\ensuremath{\PLambda}}\xspace}

\def\LorLbar     {\kern \thebaroffset\optbar{\kern -\thebaroffset \PLambda}\xspace}

\def\Lb           {{\ensuremath{\Lz^0_\bquark}}\xspace}

\newcommand{\decay}[2]{\ensuremath{#1\!\to #2}\xspace}         % Not mbox! It changes sizes

\def\to                 {\ensuremath{\rightarrow}\xspace}

\def\BsToJPsiPhi  {\decay{\Bs}{\jpsi\phi}}

\def\AT#1     {\ensuremath{A_{\mathrm{T}}^{#1}}\xspace}           % 2

\def\Bsmm     {\decay{\Bs}{\mup\mun}}

\def\C#1      {\ensuremath{\mathcal{C}_{#1}}\xspace}                       % 9
\def\Cp#1     {\ensuremath{\mathcal{C}_{#1}^{'}}\xspace}                    % 7
\def\Ceff#1   {\ensuremath{\mathcal{C}_{#1}^{\mathrm{(eff)}}}\xspace}        % 9  
\def\Cpeff#1  {\ensuremath{\mathcal{C}_{#1}^{'\mathrm{(eff)}}}\xspace}       % 7
\def\Ope#1    {\ensuremath{\mathcal{O}_{#1}}\xspace}                       % 2
\def\Opep#1   {\ensuremath{\mathcal{O}_{#1}^{'}}\xspace}                    % 7

\newcommand{\aunit}[1]{\ensuremath{\text{\,#1}}}       
\newcommand{\pev}{\aunit{Pe\kern -0.1em V}\xspace}
\newcommand{\tev}{\aunit{Te\kern -0.1em V}\xspace}
\newcommand{\gev}{\aunit{Ge\kern -0.1em V}\xspace}
\newcommand{\mev}{\aunit{Me\kern -0.1em V}\xspace}
\newcommand{\kev}{\aunit{ke\kern -0.1em V}\xspace}
\newcommand{\ev}{\aunit{e\kern -0.1em V}\xspace}
\newcommand{\mevc}{\ensuremath{\aunit{Me\kern -0.1em V\!/}c}\xspace}
\newcommand{\gevc}{\ensuremath{\aunit{Ge\kern -0.1em V\!/}c}\xspace}
\newcommand{\mevcc}{\ensuremath{\aunit{Me\kern -0.1em V\!/}c^2}\xspace}
\newcommand{\gevcc}{\ensuremath{\aunit{Ge\kern -0.1em V\!/}c^2}\xspace}
\def\fb   {\ensuremath{\aunit{fb}}\xspace}
\def\invfb   {\ensuremath{\fb^{-1}}\xspace}

\newcommand{\chisq}{\ensuremath{\chi^2}\xspace}

\def\gsim{{~\raise.15em\hbox{$>$}\kern-.85em
          \lower.35em\hbox{$\sim$}~}\xspace}
\def\lsim{{~\raise.15em\hbox{$<$}\kern-.85em
          \lower.35em\hbox{$\sim$}~}\xspace}

\def\pt         {\ensuremath{p_{\mathrm{T}}}\xspace}

\def\ptot       {\ensuremath{p}\xspace}

\def\evtgen     {\mbox{\textsc{EvtGen}}\xspace}

\def\geant      {\mbox{\textsc{Geant4}}\xspace}

\def\photos     {\mbox{\textsc{Photos}}\xspace}

\def\pythia     {\mbox{\textsc{Pythia}}\xspace}

\xspace

\def\tell1  {TELL1\xspace}
\def\ukl1   {UKL1\xspace}

%% file: our-symbols-def.tex
\newcommand{\Ry}{\ensuremath{\mathcal{R}}\xspace}

\newcommand{\mjpsikk}{\ensuremath{m(\jpsi K^+ K^-)}\xspace}
\newcommand{\mjpsik}{\ensuremath{m(\jpsi K^+)}\xspace}
\newcommand{\mjpsiphi}{\ensuremath{m(\jpsi K^+K^-)}\xspace}

\def\bujpsik{\ensuremath{\Bu \to \jpsi \Kp}\xspace}
\def\bujpsipi{\ensuremath{\Bu \to \jpsi \pi^+}\xspace}

\def\bsjpsiphi{\BsToJPsiPhi}
\def\bsjpsiphiTokSkL{\BsToJPsiPhi(\to \ensuremath{\KS \KL})\xspace}
\def\bdjpsikstarDom{\ensuremath{B^0 \to \jpsi K^{\ast}(892)^0}\xspace}
\def\bdjpsikstar{\ensuremath{B^0 \to \jpsi K^{\ast 0}}\xspace}
\def\bdjpsikpi{\ensuremath{B^0 \to \jpsi K^+ \pi^-}\xspace}

\newcommand{\Lambdab}{\ensuremath{\Lb}\xspace}

\newcommand{\BuJpsiK}{\ensuremath{B^+\to \jpsi K^+}\xspace}

\newcommand{\BsJpsiKst}{\ensuremath{B^0_s\to \jpsi \Kstarzb}\xspace}

\newcommand{\BcJpsiKKpi}{\ensuremath{B^+_c\to \jpsi K^+ K^- \pi^+}\xspace}

\newcommand{\Jpsi}{\ensuremath{\jpsi}\xspace}
\newcommand{\phimeso}{\ensuremath{\phi(1020)}\xspace}
\newcommand{\phimes}{\ensuremath{\phi}\xspace}

\newcommand{\BsJpsifTopipi}{\ensuremath{B^0_s\to \jpsi f_0 (\to \pi^+ \pi^-)}\xspace}
\newcommand{\BsJpsiPhi}{\ensuremath{B^0_s\to \jpsi \phi}\xspace}

\newcommand{\lbpjpsik}{\ensuremath{\Lambdab \to \jpsi pK^-}\xspace}

\def\pdf {PDF\xspace}
\def\pdfs {PDFs\xspace}

\newcommand{\fdfsinline}{\ensuremath{f_d/\kern -0.1emf_s}\xspace}
\newcommand{\fdfsinlinecompact}{\ensuremath{^{f_d}\kern -0.2em/\kern -0.22em_{f_s}}\xspace}

\newcommand{\fsfd}{\ensuremath{f_s/f_d}\xspace}
\newcommand{\fdfuinline}{\ensuremath{f_d/\kern -0.1emf_u}\xspace}
\newcommand{\fsfuinline}{\ensuremath{f_s/\kern -0.1emf_u}\xspace}
\newcommand{\fsfu}{\fsfuinline}

\newcommand{\figref}[1]{Fig.~\ref{#1}}
\newcommand{\tabref}[1]{Tab.~\ref{#1}}

\let\originaleqref\eqref
\renewcommand{\eqref}{Eq.~\originaleqref}

%% file: thesisstyle.tex
\def\pz         {\mbox{$p_{\mathrm{L}}$}\xspace}

\newcommand{\Bptot}{\ensuremath{{\ptot^{\B}}}\xspace}
\newcommand{\Bpz}{\ensuremath{{p_{\mathrm{L}}^{\B}}}\xspace}
\newcommand{\Bpt}{\ensuremath{{\pt^{\B}}}\xspace}
\newcommand{\Beta}{\ensuremath{{\eta^{\B}}}\xspace}
\newcommand{\By}{\ensuremath{{y^{\B}}}\xspace}
\newcommand{\deltay}{\ensuremath{{\Delta y}}\xspace}

\def\pp      {\ensuremath{pp}\xspace}

%% file: title-LHCb-PAPER.tex
% $Id: title-LHCb-PAPER.tex 122889 2018-08-17 17:59:55Z pkoppenb $
% ===============================================================================
% Purpose: LHCb-PAPER journal paper title page template
% Author: 
% Created on: 2010-09-25
% ===============================================================================

%%%%%%%%%%%%%%%%%%%%%%%%%
%%%%%  TITLE PAGE  %%%%%%
%%%%%%%%%%%%%%%%%%%%%%%%%
\begin{titlepage}
\pagenumbering{roman}

% Header ---------------------------------------------------
\vspace*{-1.5cm}
\centerline{\large EUROPEAN ORGANIZATION FOR NUCLEAR RESEARCH (CERN)}
\vspace*{1.5cm}
\noindent
\begin{tabular*}{\linewidth}{lc@{\extracolsep{\fill}}r@{\extracolsep{0pt}}}
\ifthenelse{\boolean{pdflatex}}% Logo format choice
{\vspace*{-1.5cm}\mbox{\!\!\!\includegraphics[width=.14\textwidth]{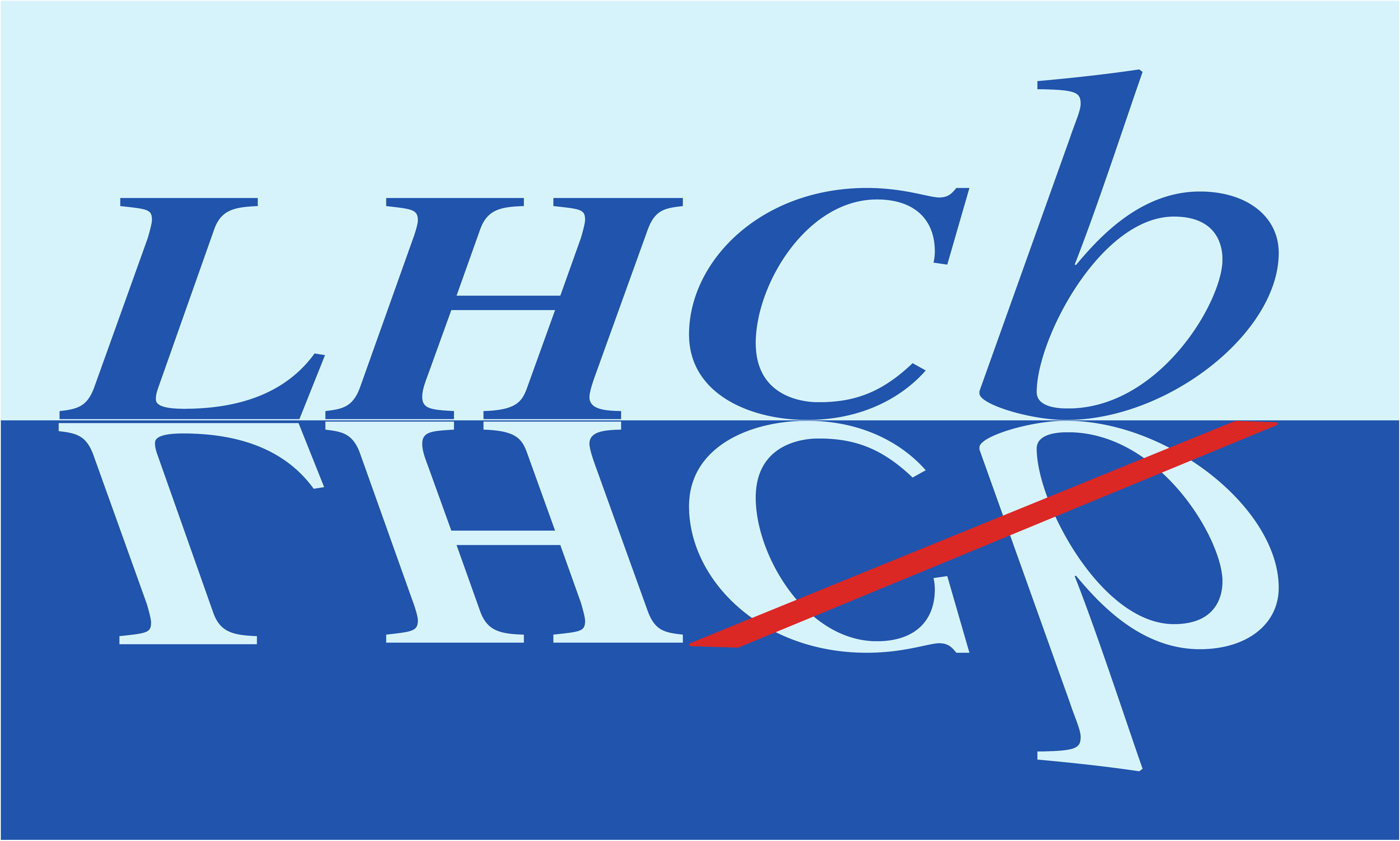}} & &}%
{\vspace*{-1.2cm}\mbox{\!\!\!\includegraphics[width=.12\textwidth]{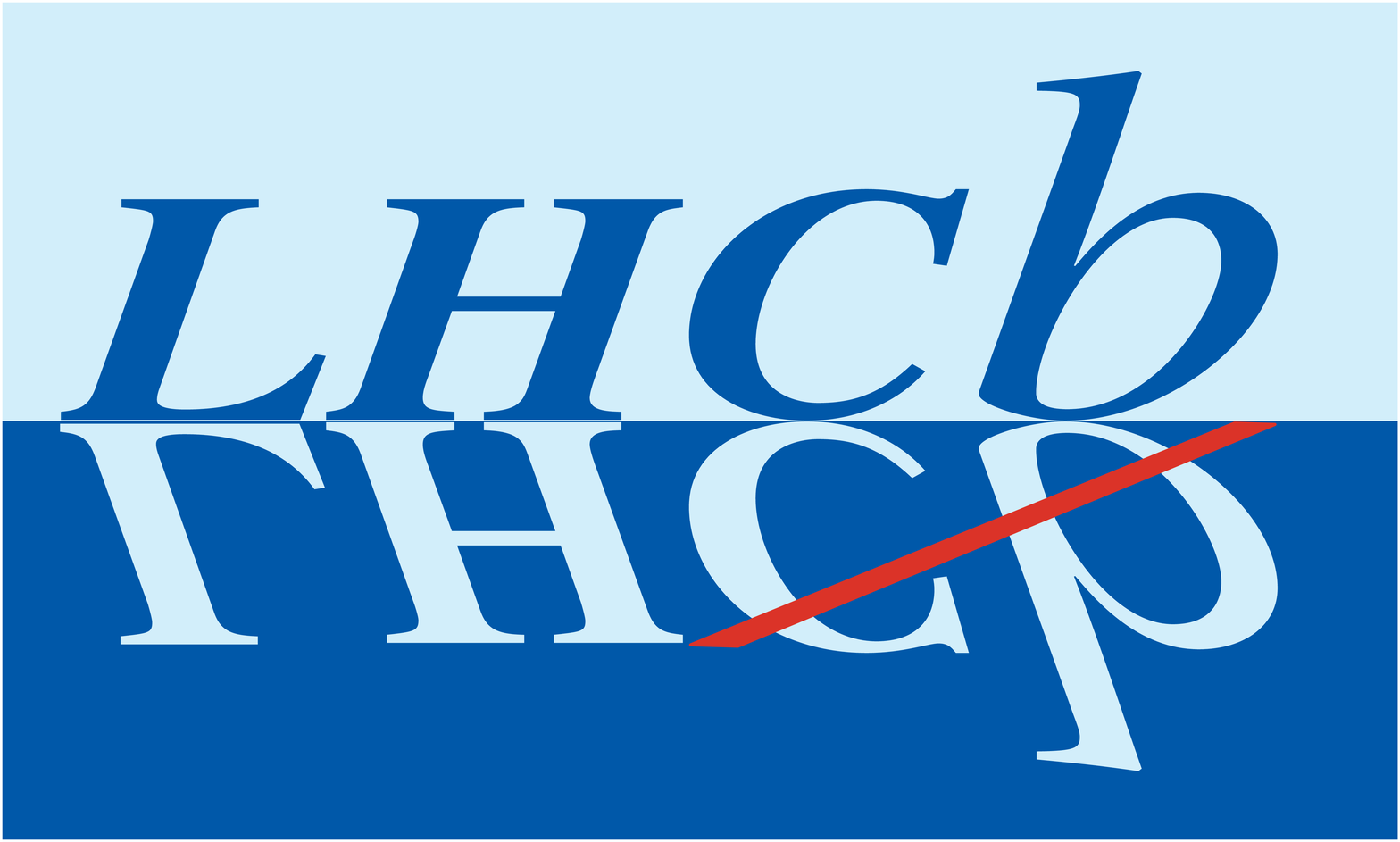}} & &}%
\\
 & & CERN-EP-2019-209 \\  % ID 
 & & LHCb-PAPER-2019-020 \\  % ID 
 & & \today \\ % Date - Can also hardwire e.g.: 23 March 2010
 & & \\
% not in paper \hline
\end{tabular*}

\vspace*{4.0cm}

% Title --------------------------------------------------
{\normalfont\bfseries\boldmath\huge
\begin{center}
% DO NOT EDIT HERE. Instead edit macro in main.tex to keep metadata correct
\papertitle 
\end{center}
}

\vspace*{2.0cm}

% Authors -------------------------------------------------
\begin{center}
%In the footnote, replace 'paper' by 'Letter' in case of submission to PRL or PLB 
% Edit macro in main.tex to keep metadata correct
\paperauthors\footnote{Authors are listed at the end of this Letter.}
\end{center}

\vspace{\fill}

% Abstract -----------------------------------------------
\begin{abstract}
The ratio of the \Bs and \Bu fragmentation fractions $f_s$ and $f_u$ is studied with \BsJpsiPhi and \BuJpsiK decays using data collected by the LHCb experiment in proton-proton collisions at 7, 8 and 13\tev center-of-mass energies. 
The analysis is performed in bins of \B-meson momentum, longitudinal momentum, transverse momentum, pseudorapidity and rapidity.
The fragmentation-fraction ratio \fsfu is observed to depend on the \B-meson transverse momentum with a significance of $6.0\,\sigma$. This dependency is driven by the 13\tev sample ($8.7\,\sigma$) 
while the results for the other collision energies are not significant when considered separately. 
Furthermore the results show a $4.8\,\sigma$ evidence for an increase of \fsfu as a function of collision energy. 
\end{abstract}

\vspace*{2.0cm}

\begin{center}
Published in Phys.~Rev.~Lett. 124 (2020) 122002
\end{center}

\vspace{\fill}

{\footnotesize 
% Edit macro in main.tex to keep metadata correct
\centerline{\copyright~\papercopyright. \href{\paperlicenceurl}{\paperlicence}.}}
\vspace*{2mm}

\end{titlepage}

%%%%%%%%%%%%%%%%%%%%%%%%%%%%%%%
%%%%%  EOD OF TITLE PAGE  %%%%%%
%%%%%%%%%%%%%%%%%%%%%%%%%%%%%%%%

%  empty page follows the title page ----
\newpage
\setcounter{page}{2}
\mbox{~}

\cleardoublepage

%% file: introduction.tex
%\section{Introduction}

The proton-proton (\pp) collisions at the LHC produce copious pairs of \bquark and \bquarkbar quarks, which immediately hadronize into the 
full spectrum of \bquark hadrons. The knowledge of $b$-hadron production rates is crucial in order to measure their branching fractions.

%The production rate of \B mesons can be described in Quantum Chromodynamics (QCD) with an integral over a product of the differential $\pp \to \bbbar$ production cross-section, \ppbbxs, 
%and a nonperturbative \emph{fragmentation function}, $\mathcal{D}_{\bquark \to B}$, as
%\ifthenelse{\boolean{wordcount}}{}{
%\begin{equation}% \label{eq:production}
%    \frac {\deriv\sigma^{B}} {\deriv\Bpt} = 
%    \int \deriv \pt^b ~ \deriv x
%    ~\frac {\deriv\ppbbxs}{\deriv \pt^b} 
%    ~\mathcal{D}_{\bquark \to B} (x) ~\delta(\Bpt-x\pt^b), % \quad,
%\end{equation}
%}
%where \Bpt ($\pt^b$) is the transverse-to-beam momentum component of the \B meson (\bquark quark) and $x$ is the fraction of $\pt^b$ carried by the \B meson~\cite{Cacciari:2001cw,Cacciari:2002pa}.
%
%Due to their nonperturbative nature, fragmentation functions for light \B mesons cannot be calculated from first principles 
%and parametric models are used instead~\cite{Mele:1990cw,Mele:1990yq,Cacciari:2001cw,Alekhin:2005dy}.
%The normalization of the integrated product is determined experimentally by measuring the \bquark-hadron \emph{fragmentation fractions} 
%in given phase-space regions.

The fragmentation fractions $f_u$, $f_d$, $f_s$, and $f_{\rm baryon}$ are defined as probabilities for a \bquark quark to hadronize into 
a \Bu, \Bd, \Bs meson or a \bquark baryon, respectively.\footnote{The inclusion of the charge-conjugate modes is implied throughout this Letter.}
These include all possible contributions from intermediate states decaying to the mentioned hadrons via strong or electromagnetic interaction. 
The \bquark-hadron fragmentation fractions were first measured in $e^+e^-$ collisions at the \Z resonance by LEP experiments~\cite{ACTON1992357,Buskulic:1995bd,Acciarri:435659,Abdallah:2003xp} and in $\proton\antiproton$ collisions at $\sqrt{s}=1.8 \tev$ center-of-mass energy by the CDF experiment~\cite{Aaltonen:2008zd}. 
In the absence of contradicting evidence the fragmentation fractions determined in different collision environments were considered 
universal and averaged~\cite{PDG2018}.

More recent measurements have shown that the hadronization fraction ratio $f_{\Lb} / f_{d}$ depends strongly on the \pt and pseudorapidity of the produced \bquark hadron~\cite{Aaltonen:2008eu,LHCb-PAPER-2014-004,LHCb-PAPER-2018-050}. 
Evidence has also been seen for a dependence on \Bpt of the relative \Bs and \Bd meson production, \fsfd~\cite{LHCb-PAPER-2012-037}.
%The kinematic dependencies have been suggested to originate from the differences in the non-perturbative hadronization part
%between the \bquark-hadron fragmentation \emph{functions}~\cite{Galanti:2015pqa,Cacciari:2002pa,Mitov:2011us}, which, 
%in combination with changes in the produced \bquark-quark spectra, could could lead to modified fragmentation \emph{fractions} at higher \pp collision energies.
%~\cite{Galanti:2015pqa,Cacciari:2002pa,Mitov:2011us}.
In combination with changes in the produced \bquark-quark spectra it could lead to modified fragmentation fraction ratios at higher \pp collision energies 
and therefore affect the branching fraction measurements which rely on normalization.
%~\cite{Galanti:2015pqa,Cacciari:2002pa,Mitov:2011us}.

This analysis studies the relative \Bs and \Bu meson production, \fsfu, dependence on \pp collision energy and on the kinematics of the produced \bquark hadron. 
Measuring the relative production is not only important for the studies of underlying QCD; \fsfu represents also an essential 
input and a dominant source of systematic uncertainty in \B branching-fraction measurements performed in hadron colliders, e.g. \Bsmm~\cite{LHCb-PAPER-2014-049,LHCb-PAPER-2017-001}. 

The analysis is performed on four independent data samples collected with the LHCb detector at three \pp collision energies: at $\sqrt{s} = 7\tev$ in the 
year 2011 (corresponding to $1\invfb$), $8\tev$ in 2012 ($2\invfb$) and at $13\tev$ in the years 2015 ($0.3\invfb$) and 2016 ($1.1\invfb$). 
The relative production of \Bs mesons to \Bu mesons in the detector acceptance is measured in each sample with the ratio of efficiency-corrected yields of \bujpsik and \bsjpsiphi decays 
\ifthenelse{\boolean{wordcount}}{}{
\begin{equation} \label{eq:effcorrat}
    \mathcal{R} \equiv \frac{ N(\bsjpsiphi) }{ N(\bujpsik)} \cdot \frac{ \epsilon(\bujpsik) } { \epsilon(\bsjpsiphi) } \propto \frac{f_s}{f_u},% \quad ,
\end{equation}}
where $\jpsi\to\mu^+\mu^-$ and $\phi \to K^+ K^-$. Here $N$ denotes the selected and reconstructed candidate yield and $\epsilon$ the related efficiency.

The study is further extended to the relative productions as a function of \B-meson kinematic variables: momentum (\Bptot), transverse momentum (\Bpt), longitudinal momentum (\Bpz), 
pseudorapidity (\Beta) and rapidity (\By).\footnote{The longitudinal momentum component is the momentum component along the beam direction.}
Due to the large uncertainty on the \bsjpsiphi branching fraction\footnote{In Ref.~\cite{Aad:2015cda} the ratio $\mathcal{R}$ was converted to an absolute \fsfd value using a theoretical prediction for the ratio of the \bsjpsiphi and \bdjpsikstar 
branching fractions~\cite{Liu:2013nea}. 
In this Letter Ref.~\cite{Liu:2013nea} is not used due to disputed theoretical uncertainties arising 
from factorization assumption.} no attempt is made to measure the absolute \fsfu value. 
In the different context of light and strange hadrons, the ALICE experiment has observed a dependence of their production ratios 
on the multiplicity of the event~\cite{ALICE:2017jyt,Acharya:2018orn,Acharya:2019kyh}. 
In this analysis this dependence is not studied, owing to technical reasons;
however such behavior will be subject of future studies.

%% file: detector.tex
%\section{Detector}

The LHCb detector~\cite{Alves:1129809,LHCb-DP-2014-002} 
is a single-arm forward spectrometer covering the (final-state track) \mbox{pseudorapidity} range $2<\eta <5$,
largely complementary to the other LHC experiments. 
The detector includes a high-precision tracking system consisting of a silicon-strip vertex detector 
surrounding the \pp interaction region, a large-area silicon-strip detector located upstream of a dipole magnet, 
three stations of silicon-strip detectors and straw drift tubes located downstream of the magnet. 
Particle identification is provided by two ring-imaging Cherenkov detectors, 
an electromagnetic and a hadronic calorimeter, and a muon system 
composed of alternating layers of iron and multi-wire proportional chambers.

%% file: selection.tex
%\section{Trigger, reconstruction and selection}

The online event selection is performed by a two-stage trigger and relies on muon candidate tracks.
The first level (hardware) trigger decision is based on information from the muon systems and selects events containing at 
least one muon with a large \pt or a pair of muons with a large product of their transverse momenta ($\sqrt{\pt\cdot\pt^{\prime}}$). 
The trigger thresholds vary between $1$ and $2 \gevc$, depending on the data-taking conditions. 

The second level (software) trigger reconstructs the full event, looks for dimuon vertices and requires them to be significantly displaced from any primary vertex (PV). 
At least one of the tracks must have $\pt > 1 \gevc$ and be inconsistent with originating from any PV. 
Only events in which the trigger decision was based on the muon tracks from the signal candidates are kept.
The muon candidates are required to pass the muon identification criteria~\cite{Archilli:2013npa}. 
No additional particle identification is required on the kaon candidates. 

Offline, the \Jpsi candidates are reconstructed by combining two oppositely charged muon tracks originating from the same vertex. 
The \phimeso candidates are reconstructed from the decays to the $K^+K^-$ final state.
The \bujpsik (\bsjpsiphi) candidates are built by combining the \Jpsi candidates with a \Kp (\phimes) candidate.
Prompt combinatorial background is suppressed by removing the events in which the \Jpsi vertex fit \chisq,
\B vertex impact-parameter or \Jpsi vertex distance, indicate that the decay vertex is either poorly reconstructed or close to the PV. 
No further selection is applied on the reconstructed \phimes vertex in order to minimize the differences between the two signal-channel selections. 
Only \Jpsi (\phimes) candidates with mass within $\pm 60\mevcc$ ($\pm 10\mevcc$) of the known \Jpsi (\phimes) masses~\cite{PDG2018} are kept;
these ranges are several times the mass resolutions of about $16\mevcc ~(3.5\mevcc)$.

Signal track candidates with momenta ${\ptot > 500\gevc}$, transverse momenta ${\pt > 40\gevc}$ 
or pseudorapidity outside of the range ${2 < \eta < 4.5}$ are removed. In addition, muon and \B transverse momenta are asked to pass ${\pt > 250\mevc}$ and ${\Bpt > 500\mevc}$ requirements, respectively.
The selected sample covers the following \B meson kinematic range:
${20 < \Bptot < 700\gevc}$, ${20 < \Bpz < 700\gevc}$, ${0.5 < \Bpt < 40\gevc}$, ${2.0 < \Beta < 6.5}$ and ${2.0 < \By < 4.5}$. 
The \Beta region between $2.0$ and $2.5$ is also accessible to the ATLAS and CMS experiments and thus important for comparison and combination of the results.

Simulated signal events are used to determine the detection efficiencies, to estimate the background contamination and to model the mass 
distributions of the selected candidates.
The simulated \pp collisions are generated using \pythia~\cite{Sjostrand:2007gs} with a specific LHCb configuration~\cite{LHCb-PROC-2010-056}. 
Hadron decays are described by \evtgen~\cite{Lange:2001uf} with final-state radiation generated using \photos~\cite{Golonka:2005pn}. 
The particle interactions with the detector material and the detector response are implemented using the \geant toolkit~\cite{Agostinelli:2002hh, LHCb-PROC-2011-006}. 
The samples of simulated signal events are corrected for known differences between data and simulation~\cite{LHCb-PAPER-2017-037}
in bins of detector occupancy and kinematic variables.
When considering the \Bs over \Bu distribution ratio, the consistency between data and simulation before correction corresponded 
to a $p$-value of at least 14\% in the kinematic variables and exceeding 90\% in the detector occupancy.

%% file: massfit.tex
%\section{Mass fit}

The signal yields are obtained by fitting the \Bu and \Bs candidate mass distributions, \mjpsik and \mjpsiphi, in the $\pm100\mevcc$ range around the known mass values using independent extended unbinned maximum-likelihood fits.
To improve the mass resolution, the \B-candidate masses are computed with the \Jpsi mass constrained to its known value~\cite{PDG2018}.

The mass distributions are described with probability density functions (\pdfs) consisting of signal, combinatorial background 
and background due to pions or protons that are wrongly identified as kaons. The signal components are parameterized by Hypatia functions~\cite{Santos:2013gra}, 
which consist of hyperbolic cores and power-law tails on both sides.
The values of the parameters that define the tails are determined from simulation. 
The combinatorial backgrounds in both models are described by exponential \pdfs.
The means and widths of the signal components and the slopes of the exponentials are unconstrained. 
The values obtained in data are larger by 10\% or less for the widths, and are consistent for the means and the other shape parameters. 
The fits repeated with fixed tails in the signal shape give consistent yield results to the constrained fits used by default. 
The contribution due to misidentified \bujpsipi decays in the \mjpsik distribution is described using a kernel density estimator technique~\cite{Cranmer:2000du} applied to simulated events. 
Its fraction, relative to the signal contribution, is found to be in agreement with the estimated fraction of $(3.8\pm 0.1)\%$.

The dominant misidentified background in the \mjpsiphi distribution arises from \bdjpsikpi decays where a pion is mistakenly reconstructed as a kaon. 
The total inclusive \bdjpsikpi background is a combination of the resonant and nonresonant contributions in the $K^+\pi^-$ final state: \bdjpsikstarDom and \bdjpsikpi.
The \pdfs of these components are linked~\cite{PhysRevD.90.112009}, each described by a combination of two Crystal Ball functions~\cite{Skwarnicki:1986xj} with a common Gaussian mean and tails on opposite sides.
The background component is included in the fit model with yield fraction defined relative to the signal contribution 
and Gaussian constrained to the expected value of ${(4.1\pm 0.5)\%}$, determined on simulation.
Contributions from the decays \BcJpsiKKpi, \BsJpsiKst, \lbpjpsik, \bsjpsiphiTokSkL and \BsJpsifTopipi are considered and found negligible.
The fit results to the \bujpsik and \bsjpsiphi candidates in 2012 data are shown in~\figref{fig:mass2012}.
Fits to all the samples are shown in the Supplemental Material (Appendix A). %at [URL will be inserted by publisher].

\begin{figure*}[t!]
    \begin{center} 
    \begin{subfigure}[t]{0.49\linewidth} 
    \begin{overpic}[width = \linewidth, trim = 0. 0. 0. 0.]{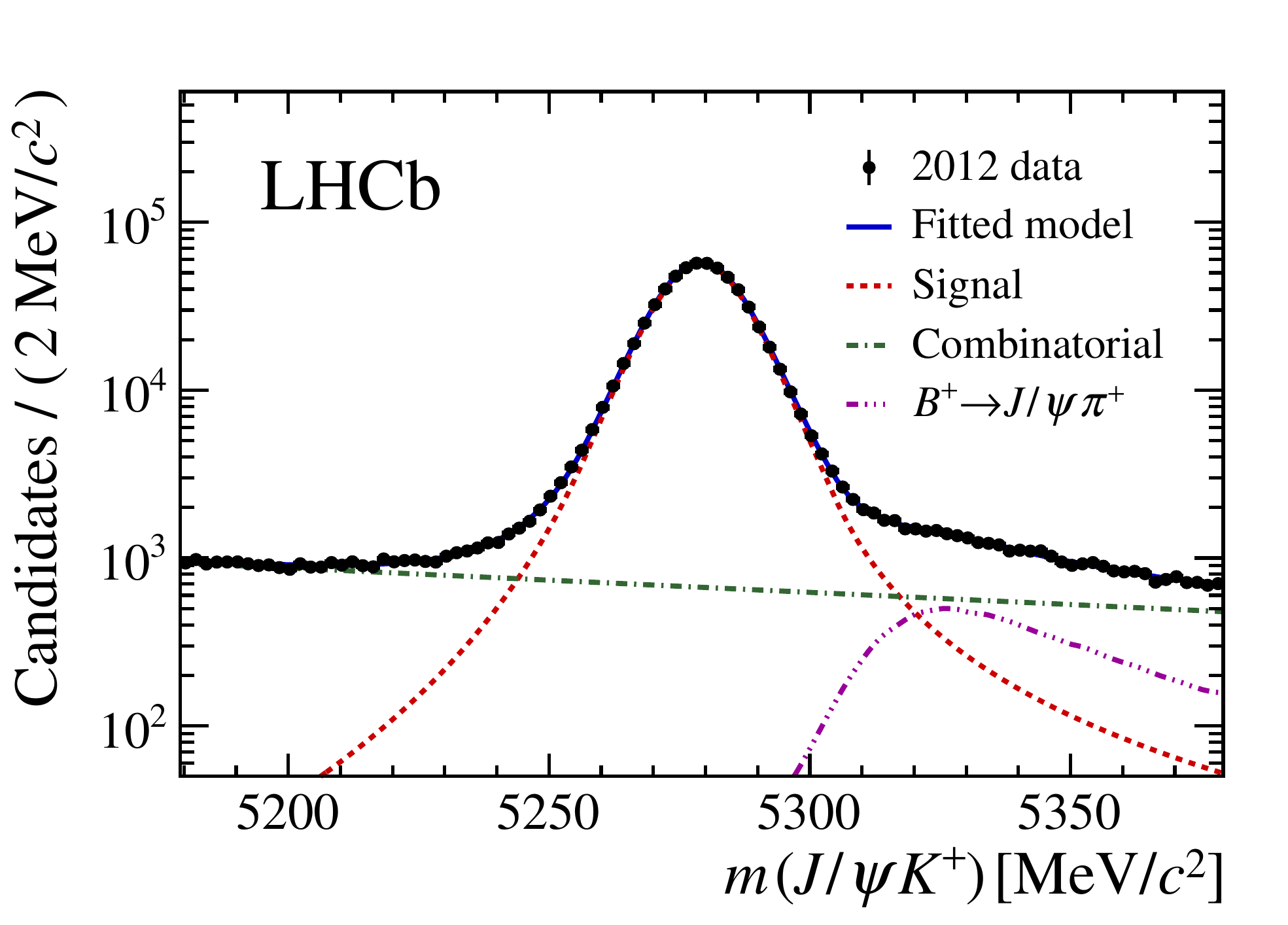}
        \put(19,50){\small (a)}
    \end{overpic}
    %\caption{\Bujpsik in 2012 data}   
    \label{fig:bujpsik_12}
    \end{subfigure}          
    \begin{subfigure}[t]{0.49\linewidth} 
    \begin{overpic}[width = \linewidth, trim = 0. 0. 0. 0.]{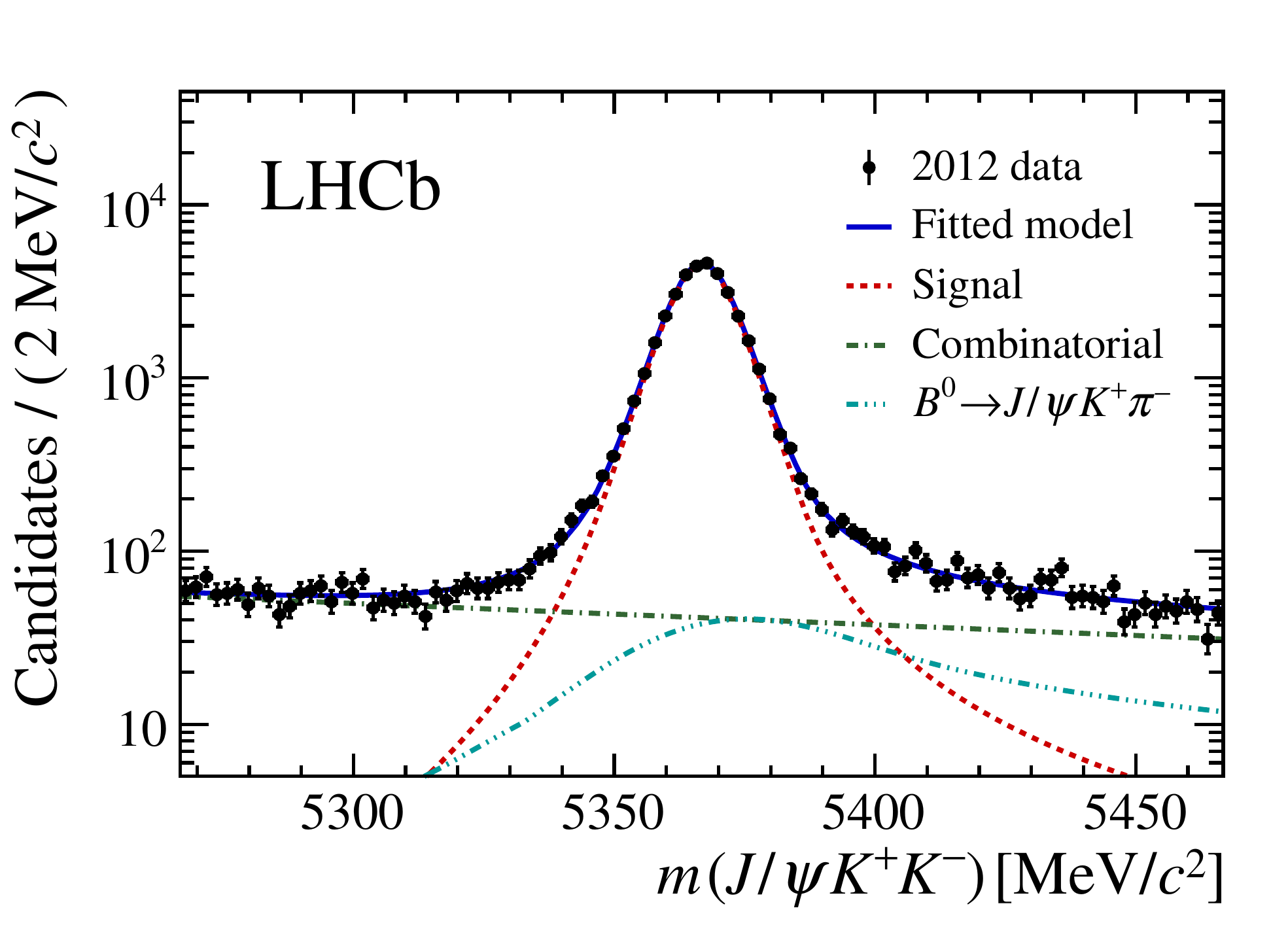}
        \put(19,50){\small (b)}
    \end{overpic}
    %\caption{\Bs in 2012 data}   \label{fig:bsjpsiphi_12}
    \end{subfigure}
    \caption{\small Mass distributions of (a) \BuJpsiK  and (b) \BsJpsiPhi candidates in the 2012 data. The result of the fit is drawn with a blue solid line. The model components are denoted with red dashed line for the signal,
    green dot-dashed line for the combinatorial background,
    magenta triple-dot-dashed line for misidentified \bujpsipi and cyan triple-dot-dashed line for misidentified the inclusive \bdjpsikpi contribution.}
    \label{fig:mass2012}
    \end{center} 
\end{figure*}

%% file: efficiencies.tex
%\section{Efficiencies}

The signal detection efficiencies include the detector acceptance, reconstruction 
efficiencies and the selection efficiencies. 
The efficiencies are computed using simulated samples unless stated otherwise. 
Tracking efficiency differences in data and simulation are corrected for. The corrections are applied for each final-state track separately in bins of the track \pt and $\eta$, and event multiplicity~\cite{Aaij:2014pwa}.

Trigger efficiencies are determined on data, separately for each data sample~\cite{Tolk:1701134}.
The trigger decision in every event can be ascribed to the reconstructed signal candidate and/or the rest of the event. 
The trigger efficiency is measured through the overlap of the two categories~\cite{LHCb-DP-2012-004}. 
The abundant \bujpsik sample is used to build a two-dimensional trigger efficiency map as a function of the \pt and \pz of the \jpsi candidates.
The choice of variables accounts for small differences in the \jpsi kinematic distributions from \bujpsik and \bsjpsiphi decays.
The average signal trigger efficiencies are computed by weighting the map contents with the fractions of simulated events in each bin and averaging the results, separately for each signal mode.
In case of the results in \B meson kinematic bins, the trigger efficiency maps are defined in bins of the considered kinematic variable 
and of an independent variable: \pt of the \jpsi candidate for the \fsfu results 
as function of \Beta, \Bpz and \By, and the \pz of the \jpsi candidate for results as a function of \Bpt.

%% file: syst.tex
%\section{Systematic uncertainties}

 Identical trigger selection and near-identical reconstruction and offline selection significantly reduce the uncertainties 
affecting the efficiency corrected \bsjpsiphi and \bujpsik yield ratio measurement.
Due to the similarity of \jpsi kinematic distributions from \bujpsik and \bsjpsiphi decays, the
efficiency ratios are close to unity, being about 0.98 for acceptance and selection and 0.99 for the trigger. 
The systematic uncertainties associated with acceptance, reconstruction and selection efficiency arise only from the limited size of simulated samples.
The dominant systematic uncertainties arise from the track-reconstruction efficiency corrections and the fit. 
A systematic uncertainty of $0.4\%$ ($0.8\%$) is assigned, following the procedures in Ref.~\cite{LHCb-DP-2013-002}, 
to the extra kaon track in \BsJpsiPhi decays in 2011 and 2012 (2015 and 2016) samples. 
For all the samples, the uncertainty is increased by an additional $1.1\%$ due to the interactions between the hadrons and detector material~\cite{LHCb-DP-2013-002}. 

The systematic uncertainty arising from the fit model is propagated to the fitted signal yields by allowing the parameters to float within Gaussian constraints with mean and width determined from the simulation.
Most of the signal and misidentified background component shape parameters are constrained with the remaining (partially correlated) tail parameters fixed to the values
determined from simulation.
The effect of fixing or leaving the signal parameters free has a negligible effect on the yield.

The resonant and nonresonant structure of the \mjpsikk spectrum is measured in Ref.~\cite{LHCb-PAPER-2012-040}. 
The resonant $f_0(980)$ meson contribution, nonresonant $S$-wave contribution and the interference effects are studied on simulated samples. No attempt is made to separate these contributions from the signal decays and the uncertainty of the fitted inclusive \bsjpsiphi yield is increased by $0.8\%$, relative to the yield.

The fit models are validated using the fitted \pdfs to generate and fit a large number of simulated pseudoexperiments according to the observed candidate yields. The pseudoexperiments are generated for the fits on the full samples as well as for the fits in bins of \Bpt and \Beta.
The mass fits in the \Bpt and \Beta bins do not show a significant bias and no additional systematic uncertainty is included. 
The pseudoexperiments for the full samples show a small yield estimator bias, the largest of which is 20\% of the statistical uncertainty. 
The uncertainties on these yields are therefore increased by the same amount to account for this.

The validity of the mass models over the \B-meson phase space is verified by comparing the fitted fractions and the model parameters across the samples and bins.
The \bujpsik fit is performed with the \bujpsipi background shape determined independently in high- and low-\Bpt regions of the simulated decays.
The variation in the observed yield is negligible. The background shapes in regions of \Beta are very similar. 
The misidentified \bdjpsikpi background \pdf variation in \Bpt or \Beta regions is studied with simulation.
The distributions show no evidence for significant variation and no additional uncertainty is assigned to the fits in bins due to the assumption of the same fit model.

%% file: results.tex
%\section{Results}

\ifthenelse{\boolean{prlversion}}%
{ \onecolumngrid
}{}
\begin{table}[bt]
    \footnotesize
    \begin{center} 
    \caption{\small 
        Efficiency-corrected \bsjpsiphi and \bujpsik yield ratios (\Ry) and uncertainties ($\sigma_{\rm{tot}}$), including the
        statistical uncertainty ($\sigma_{\rm{stat}}$) and
        the fully correlated and uncorrelated systematic uncertainties among the samples ($\sigma^{\rm{uncor}}_{\rm{syst}}$, $\sigma^{\rm{cor}}_{\rm{syst}}$). Individual contributions from tracking efficiency ($\sigma^{\rm{track}}_{\rm{syst}}$), acceptance, reconstruction and selection efficiency ($\sigma^{\rm{sel}}_{\rm{syst}}$) and fit model ($\sigma^{\rm{fit}}_{\rm{syst}}$) are shown separately. Correlations stem from the common tracking and fit model uncertainties.
        }
    %\hspace*{-2.4cm}
\ifthenelse{\boolean{wordcount}}{}{
    \begin{tabular}{lccc|c|cc|ccc}
    \toprule
    Year    &$\sqrt{s}$    &   \Ry    & $\sigma_{\rm{tot}}$      & $\sigma_{\rm{stat}}$ &  $\sigma_{\rm{syst}}^{\rm{uncor}}$  & 
    $\sigma^{\rm{cor}}_{\rm{syst}}$  &    $\sigma^{\rm{track}}_{\rm{syst}}$     &   $\sigma^{\rm{sel}}_{\rm{syst}}$    &   $\sigma^{\rm{fit}}_{\rm{syst}}$   \\
    \midrule
    2011        & \phantom{0}7\tev      &  0.1238       &  $0.0024$     &         $0.0010$                & $0.0018$                  & $0.0012$               & $0.0015$ & $0.0008$ & $0.0013$   \\
    2012        & \phantom{0}8\tev      &  0.1270       &  $0.0023$     &        $0.0007$                & $0.0019$                  & $0.0012$               & $0.0016$ & $0.0005$ & $0.0015$   \\
    2015        &13\tev                 &  0.1338       &  $0.0030$     &       $0.0017$                & $0.0022$                  & $0.0012$               & $0.0019$ & $0.0004$ & $0.0016$   \\
    2016        &13\tev                 &  0.1319       &  $0.0024$     &      $0.0008$                & $0.0021$                  & $0.0007$               & $0.0018$ & $0.0004$ & $0.0012$ \\
    \bottomrule
    \label{tab:ratios}
    \end{tabular}}
    \end{center} 
\end{table}

\ifthenelse{\boolean{prlversion}}%
{ \twocolumngrid
}{}

\begin{figure}[tb!]
    \begin{center} 
    \ifthenelse{\boolean{prlversion}}
            {\includegraphics[width = \linewidth, trim = 0. 0. 0. 0.]{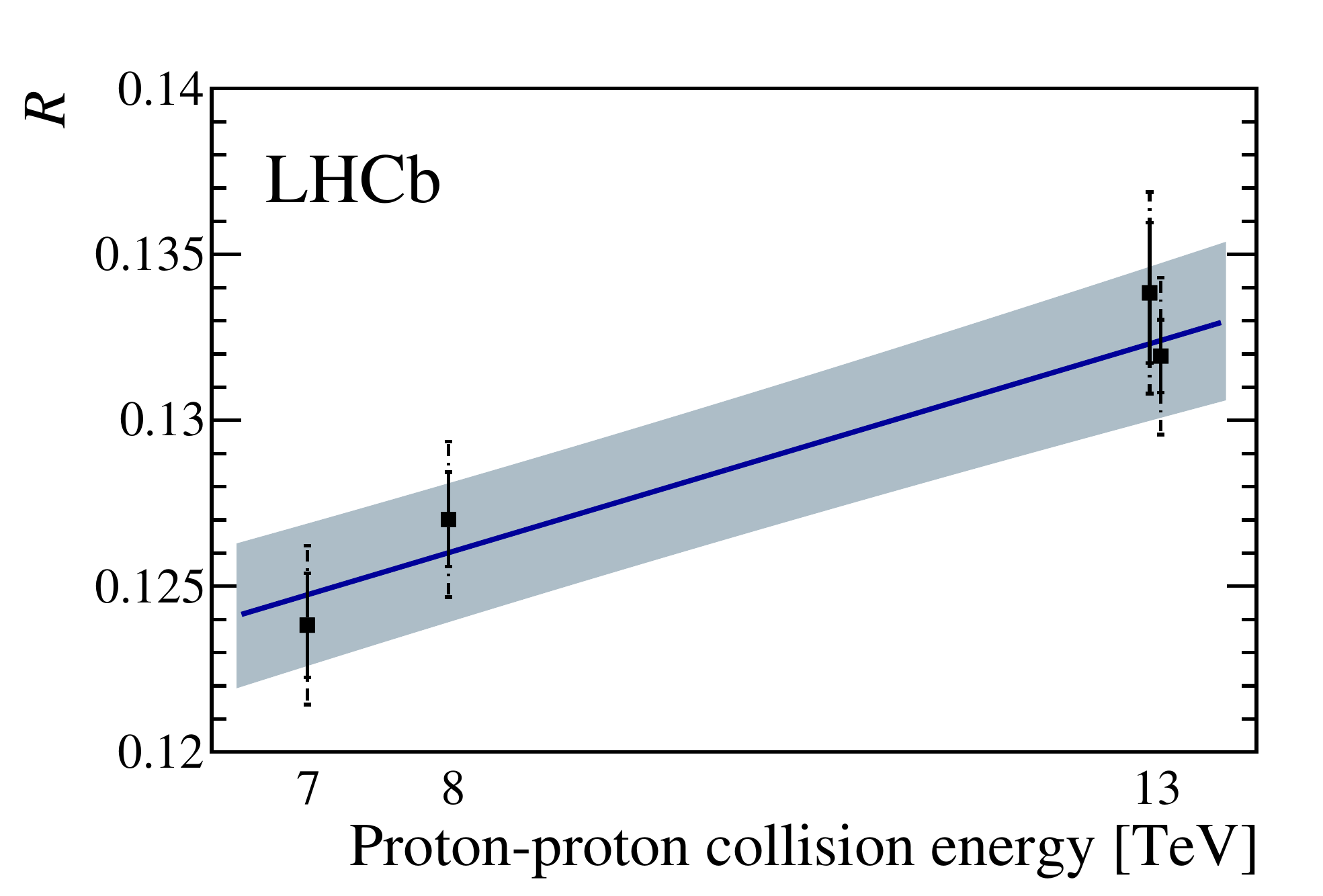} }
            {\includegraphics[width = 0.7\linewidth, trim = 0. 0. 0. 0.]{Fig2.pdf} }
    \caption{\small 
        Efficiency-corrected $\protect \bsjpsiphi$ and $\protect\bujpsik$ yield ratios (\Ry) at different $\protect \pp$ collision energies with the total (uncorrelated, including statistical) uncertainties denoted by dashed (solid) error bars. The fit result is shown with the blue solid line, the blue band denotes the $\protect 68\%$ confidence region. The 13\tev measurements are shifted horizontally for clarity.}
        \label{fig:energy}
    \end{center} 
\end{figure}

The ratios ($\mathcal{R}$) and their detailed uncertainty composition are shown in~\tabref{tab:ratios}. The ratios are fitted as a function of the \pp collision energy with a two-parameter function: $a + k_s  \sqrt{s}$,
as shown in~\figref{fig:energy}. The statistical significance of the \fsfu dependence on collision energy is estimated by comparing this fit with that under the null hypothesis $k_s=0$.  The $\chisq$ difference between the two cases is used as a test statistic and its $p$-value is determined from the $\chisq$ distribution with one degree of freedom~\cite{Wilks:1938dza}.
The two-sided significance of the two-parameter fit ($a = 0.1159\pm 0.0032$, $k_s = (1.27\pm 0.27) \times 10^{-3} \tev^{-1}$, correlation $\rho = -0.76$) is $4.8\,\sigma$ with respect to the hypothesis of no energy dependence. 
The fit accounts for the correlations between the samples due to the common tracking and fit uncertainties as described in Appendix~\ref{sec:Supplementary-App}.

The measured double ratios for different collision energies are 
\ifthenelse{\boolean{wordcount}}{}{
\begin{align}
   \mathcal{R}_{8\tev} / \mathcal{R}_{7\tev}   &= 1.026 \pm 0.017 , \nonumber \\
    \mathcal{R}_{13\tev} / \mathcal{R}_{7\tev}   &= 1.068 \pm 0.016 , \nonumber
    \label{eq:DoubRat}
\end{align}}
with the correlation coefficient $\rho = 0.33$ between the two and the correlated uncertainties accounted for.

\begin{figure*}[t]
    \ifthenelse{\boolean{prlversion}}
    {
    \begin{subfigure}[t]{0.32\linewidth} 
    \begin{overpic}[width = \linewidth, trim = 1cm 0. 1cm. 0.]{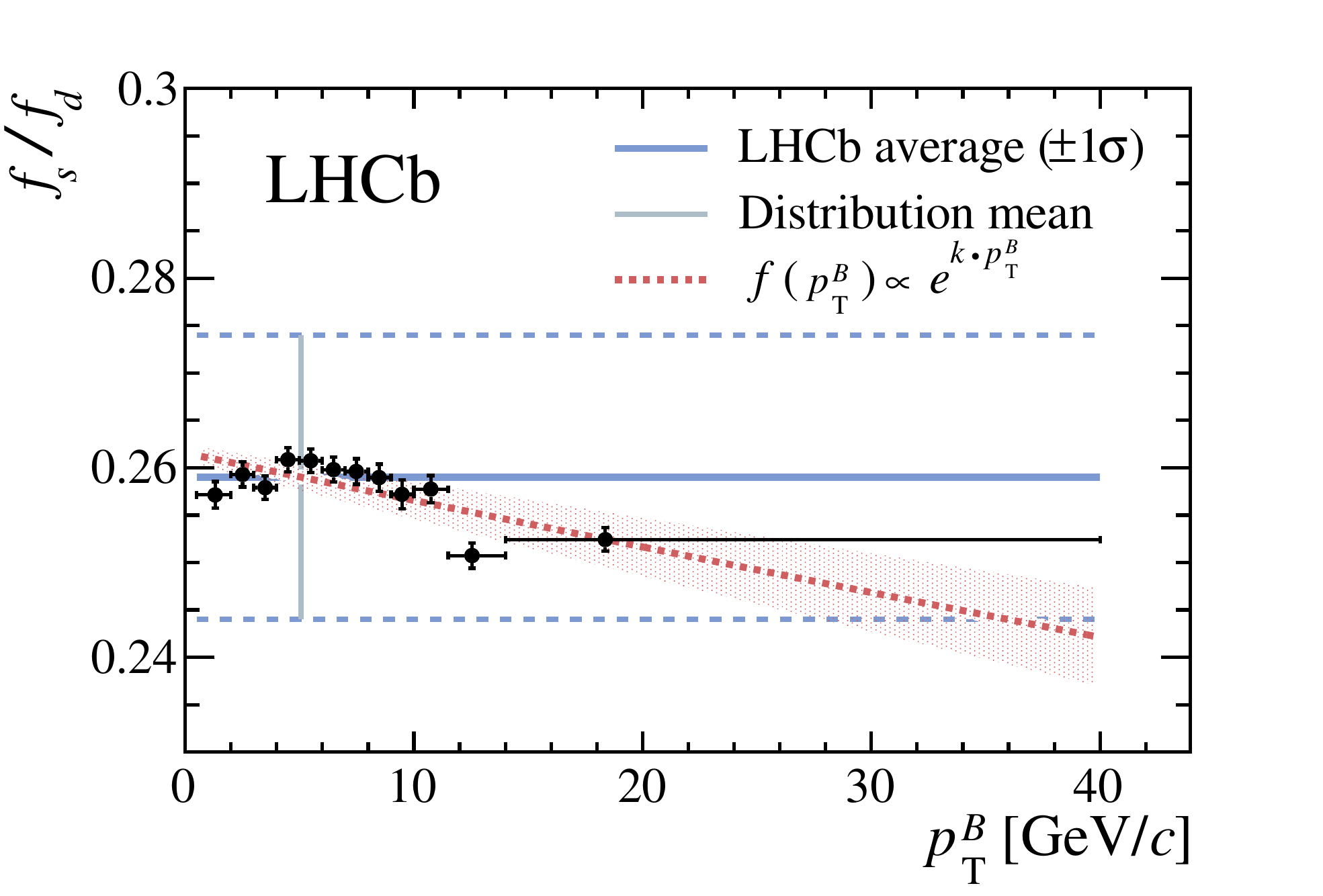}
        \put(16,52){\small (a)}
    \end{overpic}
    \end{subfigure} 
    \begin{subfigure}[t]{0.32\linewidth} 
    \begin{overpic}[width = \linewidth, trim = 1cm 0. 1cm. 0.]{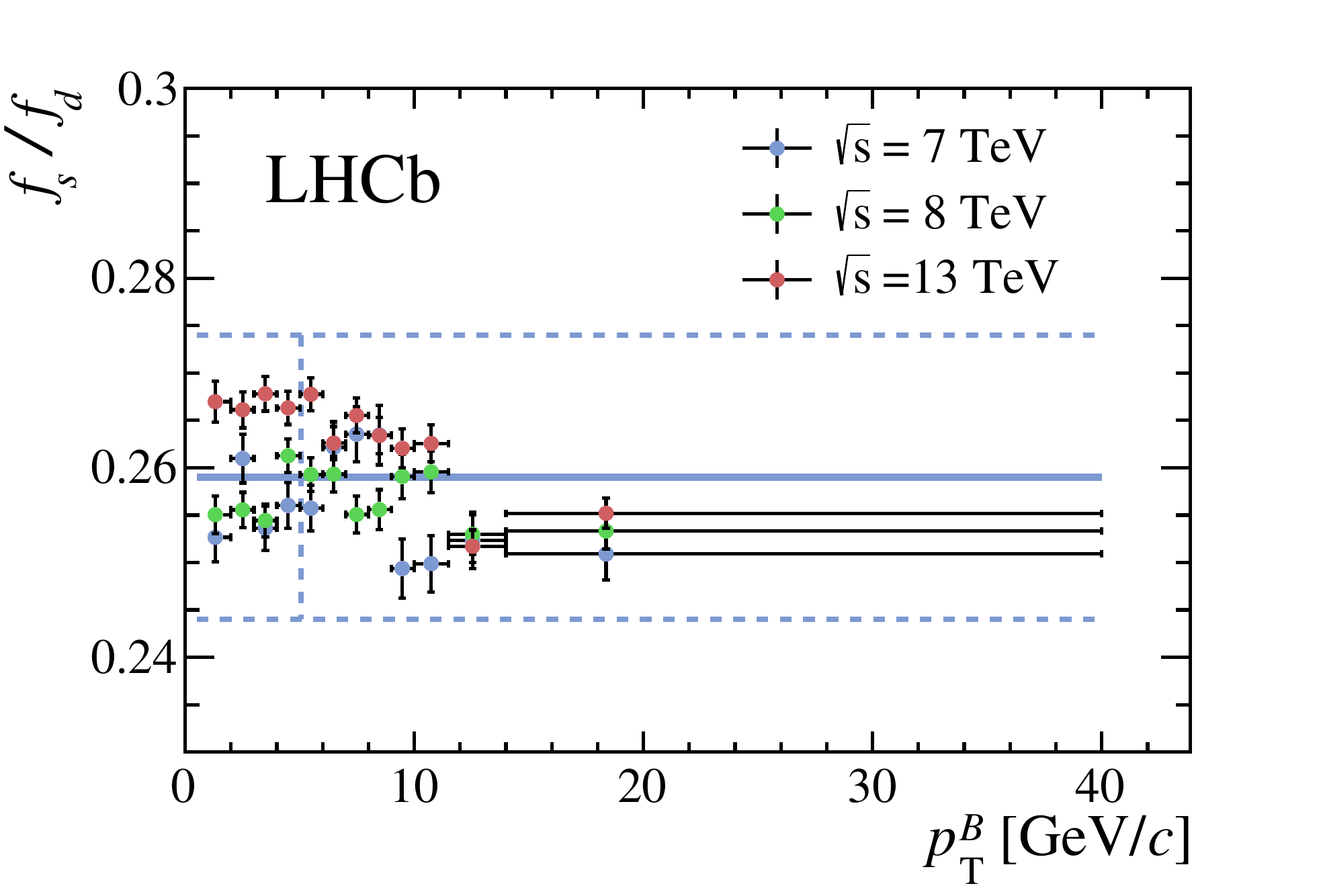} 
        \put(16,52){\small (b)}
    \end{overpic}
    \end{subfigure}
    \begin{subfigure}[t]{0.32\linewidth} 
    \begin{overpic}[width = \linewidth, trim = 1cm 0. 1cm 0.]{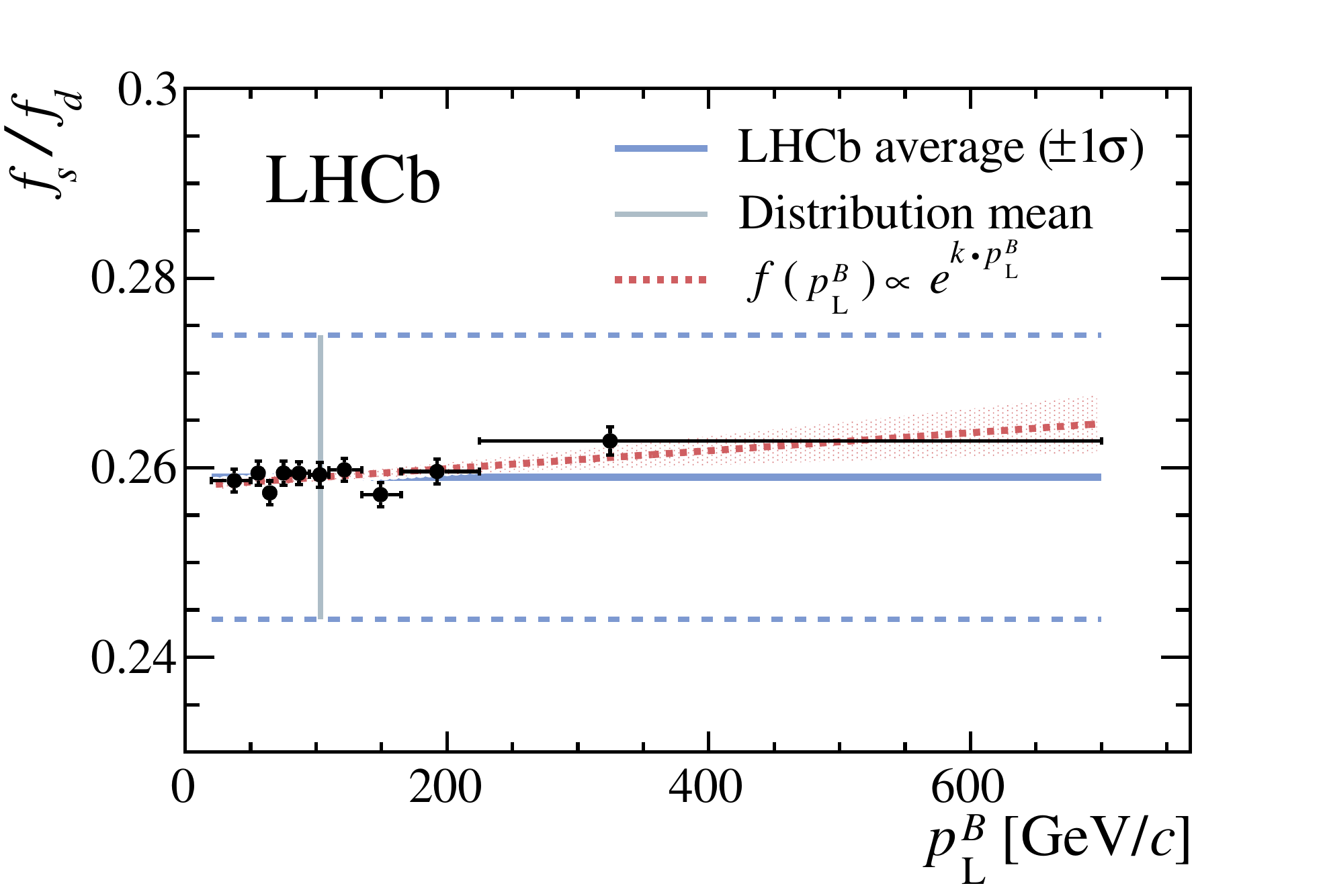}
        \put(16,52){\small (c)}
    \end{overpic}
    \end{subfigure}
    \begin{subfigure}[t]{0.32\linewidth} 
    \begin{overpic}[width = \linewidth, trim = 1cm 0. 1cm 0.]{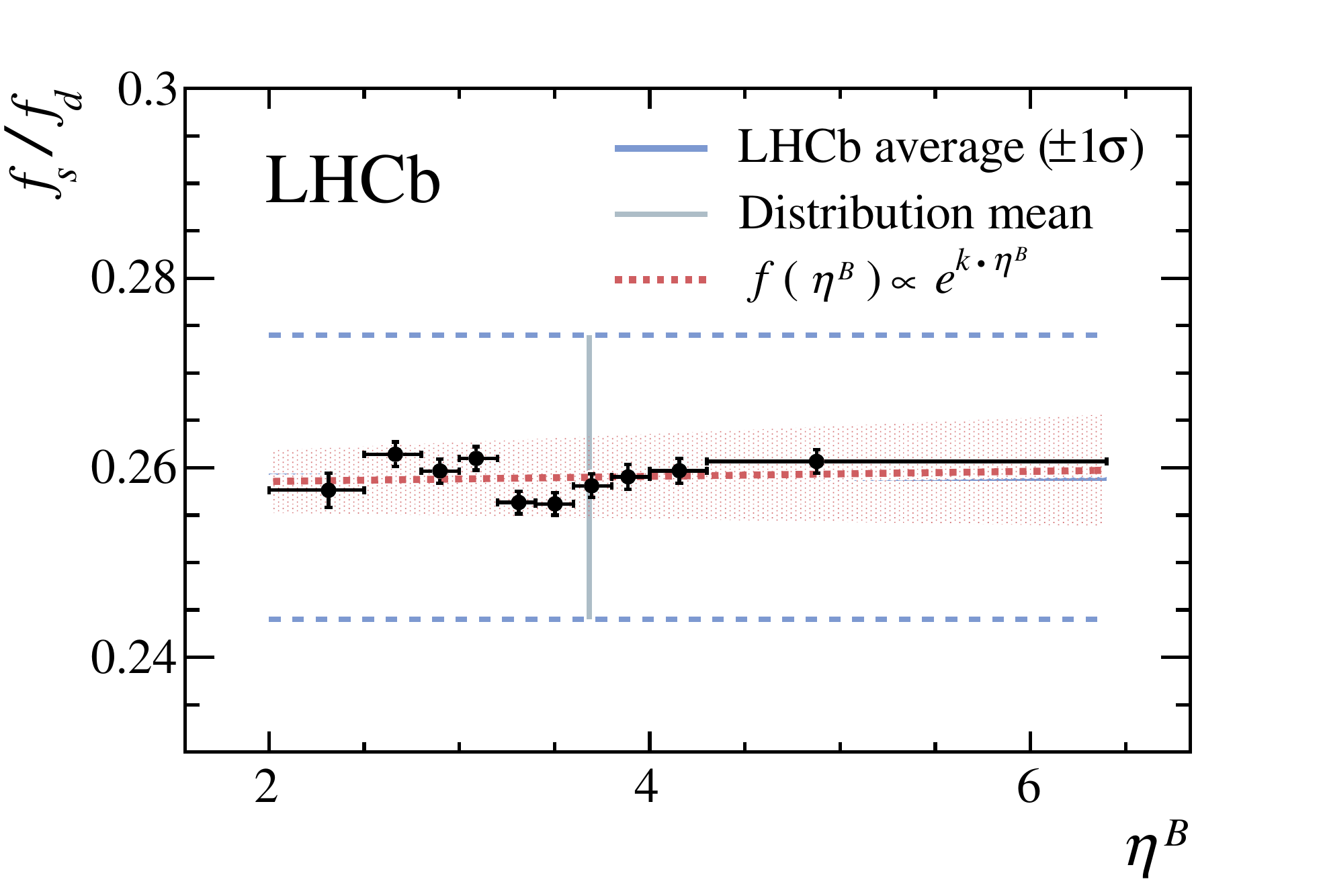}
        \put(16,52){\small (d)}
    \end{overpic}
    \end{subfigure}
    }{
    \begin{center} 
    \begin{subfigure}[t]{0.49\linewidth} 
    \begin{overpic}[width = \linewidth, trim = 1cm 0. 1cm. 0.]{Fig3_a.pdf}
    \put(16,52){\small (a)}
    \end{overpic}
    \end{subfigure}        
    \begin{subfigure}[t]{0.49\linewidth} 
    \begin{overpic}[width = \linewidth, trim = 1cm 0. 1cm. 0.]{Fig3_d.pdf} 
        \put(16,52){\small (b)}
    \end{overpic}
    \end{subfigure}
    \begin{subfigure}[t]{0.49\linewidth} 
    \begin{overpic}[width = \linewidth, trim = 1cm 0. 1cm 0.]{Fig3_b.pdf}
    \put(16,52){\small (c)}
    \end{overpic}
    \end{subfigure}
    \begin{subfigure}[t]{0.49\linewidth} 
    \begin{overpic}[width = \linewidth, trim = 1cm 0. 1cm 0.]{Fig3_c.pdf}
    \put(16,52){\small (d)}
    \end{overpic}
    \end{subfigure}
    \end{center} 
} 

    \ifthenelse{\boolean{prlversion}}
    {
    \caption{\small
    Efficiency-corrected \bsjpsiphi and \bujpsik yield ratios (\Ry) in bins of (a)  \Bpt, 
    (c) \Bpz and (d) \Beta. The ratios are scaled 
    to match the measured \fsfd value (horizontal blue lines, the $\pm1\sigma$ interval is indicated by the dashed blue lines) at the positions indicated by the vertical gray lines. 
    The red dashed lines denote the results of the exponential fits used to estimate the statistical significances of the variations (see text), and the red band denotes the $\protect 68\%$ confidence region.
    In (b) the results as a function of \Bpt are obtained separately in the three collision energies. }
    } 
    {
    \caption{\small
    Efficiency-corrected \bsjpsiphi and \bujpsik yield ratios (\Ry) in bins of (a)  \Bpt, 
    (c)  \Bpz and (d) \Beta. The ratios are scaled 
    to match the measured \fsfd value (horizontal blue lines, the $\pm1\sigma$ interval is indicated by the dashed blue lines) at the positions indicated by the vertical gray lines. 
    The red dashed lines denote the results of the exponential fits used to estimate the statistical significances of the variations (see text).
    In (b) the results as a function of \Bpt are obtained separately in the three collision energies. }
    }
    \label{fig:light}
    %\end{center} 
\end{figure*}

In each sample, the efficiency-corrected signal yield ratios are measured in bins of the \B-meson kinematic variables $v \in \{\Bptot, \Bpt, \Bpz, \Beta, \By\}$ and averaged.
On the vertical scale of~\figref{fig:light}, the averaged signal-yield ratios are scaled, assuming $f_u = f_d$, to match the average \fsfd value measured at $\sqrt{s} = 7\tev$ 
($\fsfd = 0.259$)~\cite{LHCb-PAPER-2011-018,LHCb-PAPER-2012-037,LHCb-CONF-2013-011} at the corresponding variable distribution means; this is for illustrative purpose alone.
On the horizontal scale, each data point is set to the mean value determined from simulation.
The statistical significance of the \fsfu dependence is estimated by fitting the $\mathcal{R}$ distributions with a function $A_v\cdot\exp(k_v\cdot v)$ under two hypotheses: 
one where no variation is allowed and the slope parameter, $k_v$, is fixed to zero and one with $k_v$ left free.

The relative \Bs and \Bu production is observed to depend on the \Bpt with a significance of $6\,\sigma$ and the fitted slope parameter is 
$k_{\Bpt}= -(1.93\pm0.46)\times10^{-3} \gev^{-1}c$. 
The strongest variation is measured for the $13\tev$ samples: $8.7\,\sigma$, ${k_{\Bpt}= -(4.40\pm0.67)\times10^{-3} \gev^{-1}c}$,
while is not significant ($2.1\,\sigma$ and $1.5\,\sigma$) for the 7 and 8 \tev results obtained separately; see Appendix A for further details. 
The variation in \Bpt is further studied in three subregions of \Bpz (${[20, 75, 125, 700]\gevc}$) and a clear dependence is seen in all the regions.
The results for \Bpt, \Bpz and \Beta are shown in~\figref{fig:light}. 
No evidence is found for significant \fsfu variation in \Bptot, \Bpz, \Beta or \By. 
For the numerical results in all the studied variables and additional figures see Appendix A.

In conclusion, the \Bs and \Bu fragmentation fraction ratio \fsfu is studied at $7\tev$, $8\tev$, and $13\tev$ \pp collision energies and in different \B-meson kinematic regions.
A $4.8\,\sigma$ evidence is seen for an \fsfu dependence on the collision energy and \fsfu is observed to depend on the \B-meson transverse momentum. 
The observed \Bpt dependence is compatible with the recent LHCb result on semileptonic modes~\cite{LHCb-PAPER-2018-050}.
No evidence of \fsfu variation is seen in \B-meson momentum, longitudinal momentum, rapidity or pseudorapidity.

%% file: acknowledgements.tex
\section*{Acknowledgements}
%
% These Acknowledgements valid from 3-May-2019
%
\noindent We express our gratitude to our colleagues in the CERN
accelerator departments for the excellent performance of the LHC. We
thank the technical and administrative staff at the LHCb
institutes.
We acknowledge support from CERN and from the national agencies:
CAPES, CNPq, FAPERJ and FINEP (Brazil); 
MOST and NSFC (China); 
CNRS/IN2P3 (France); 
BMBF, DFG and MPG (Germany); 
INFN (Italy); 
NWO (Netherlands); 
MNiSW and NCN (Poland); 
MEN/IFA (Romania); 
MSHE (Russia); 
MinECo (Spain); 
SNSF and SER (Switzerland); 
NASU (Ukraine); 
STFC (United Kingdom); 
DOE NP and NSF (USA).
We acknowledge the computing resources that are provided by CERN, IN2P3
(France), KIT and DESY (Germany), INFN (Italy), SURF (Netherlands),
PIC (Spain), GridPP (United Kingdom), RRCKI and Yandex
LLC (Russia), CSCS (Switzerland), IFIN-HH (Romania), CBPF (Brazil),
PL-GRID (Poland) and OSC (USA).
We are indebted to the communities behind the multiple open-source
software packages on which we depend.
Individual groups or members have received support from
AvH Foundation (Germany);
EPLANET, Marie Sk\l{}odowska-Curie Actions and ERC (European Union);
ANR, Labex P2IO and OCEVU, and R\'{e}gion Auvergne-Rh\^{o}ne-Alpes (France);
Key Research Program of Frontier Sciences of CAS, CAS PIFI, and the Thousand Talents Program (China);
RFBR, RSF and Yandex LLC (Russia);
GVA, XuntaGal and GENCAT (Spain);
the Royal Society
and the Leverhulme Trust (United Kingdom).

%% file: supplementary-app.tex
\clearpage

\section{Supplementary material for LHCb-PAPER-2019-020}
\label{sec:Supplementary-App}

\subsection{\boldmath Details on the energy variation fit}

The four ratios of efficiency corrected \BsJpsiPhi and \BuJpsiK yields (\tabref{tab:ratios}) are fitted with a linear function: $a+k_s\cdot \sqrt{s}$. 
The ratios are correlated due to the common tracking efficiency systematic uncertainties and due to the common ($0.8\%$) systematic uncertainty 
assigned to the fitted \BsJpsiPhi yield in order to account for additional resonant and nonresonant contributions.
The following covariance matrix is used to account for the correlations in the $\chi^2$ fit:
 \begin{align*}
       \begin{pmatrix}
           5.737       &   3.351  &   3.948 &  3.790 \\
           3.351       &   5.471  &   4.049 &  3.886 \\
           3.948       &   4.049  &   9.219 &  4.579 \\
           3.790        &   3.886 &   4.579 &  5.595
       \end{pmatrix} \times 10^{-6}.
   \end{align*}

\clearpage

\subsection{\boldmath Fitted \boldmath\B-meson mass distributions}

\begin{figure}[!h]
    \begin{center} 
    \begin{subfigure}[t]{0.40\linewidth} 
    \includegraphics[width = \linewidth, trim = 0. 0. 0. 0.]{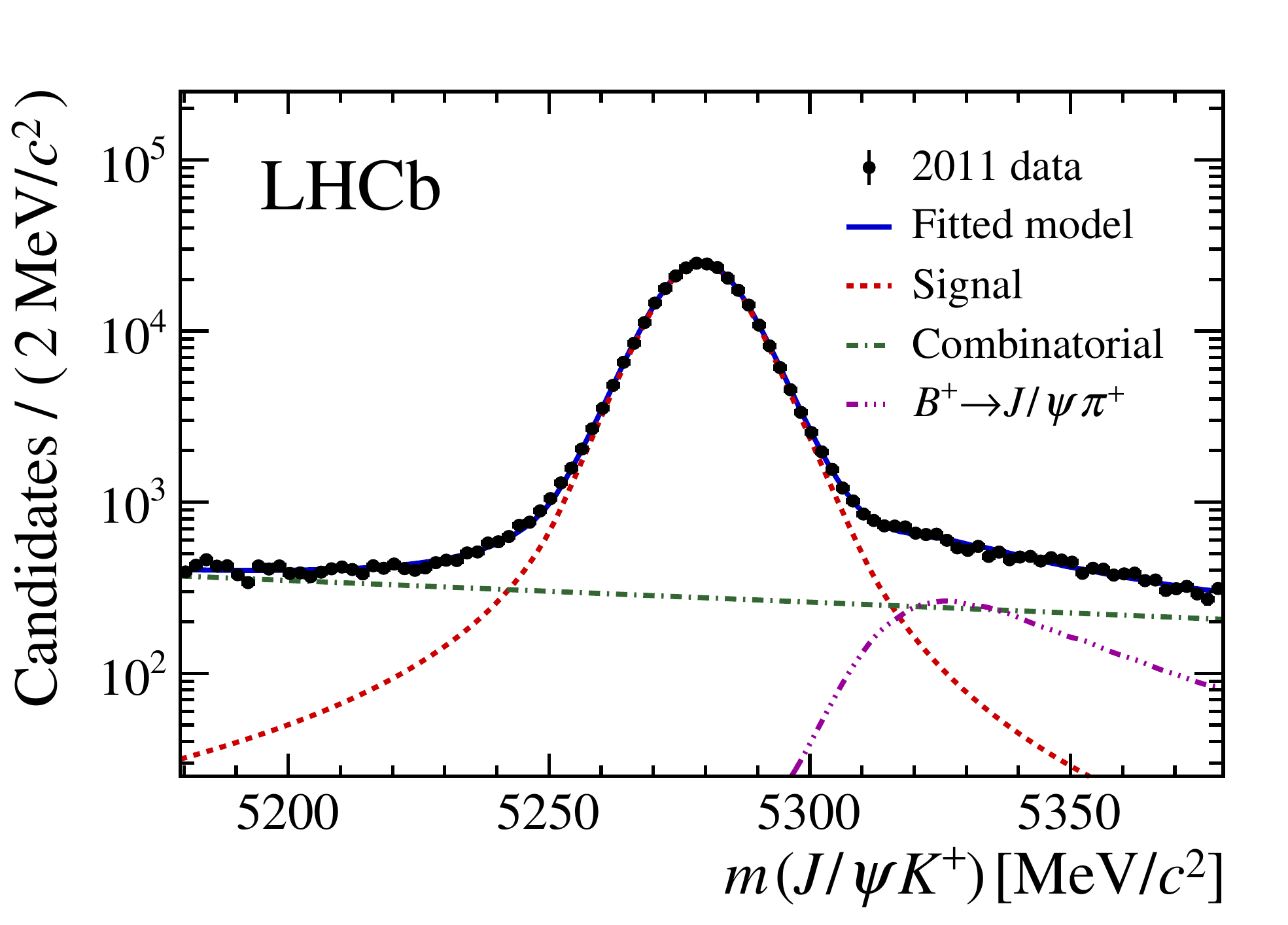}
    \end{subfigure}          
    \begin{subfigure}[t]{0.40\linewidth} 
    \includegraphics[width = \linewidth, trim = 0. 0. 0. 0.]{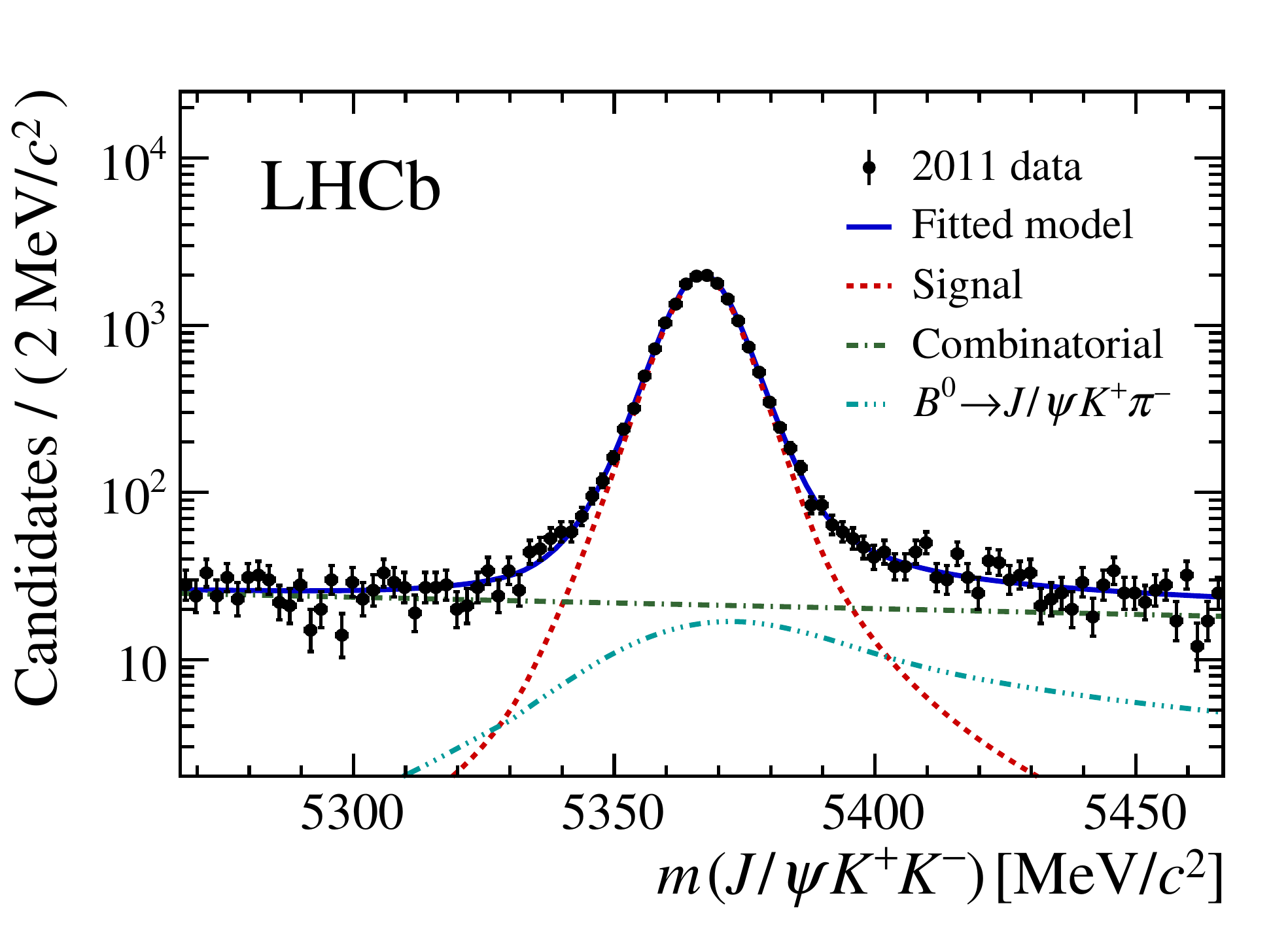}
    \end{subfigure}\\
     \begin{subfigure}[t]{0.40\linewidth} 
    \includegraphics[width = \linewidth, trim = 0. 0. 0. 0.]{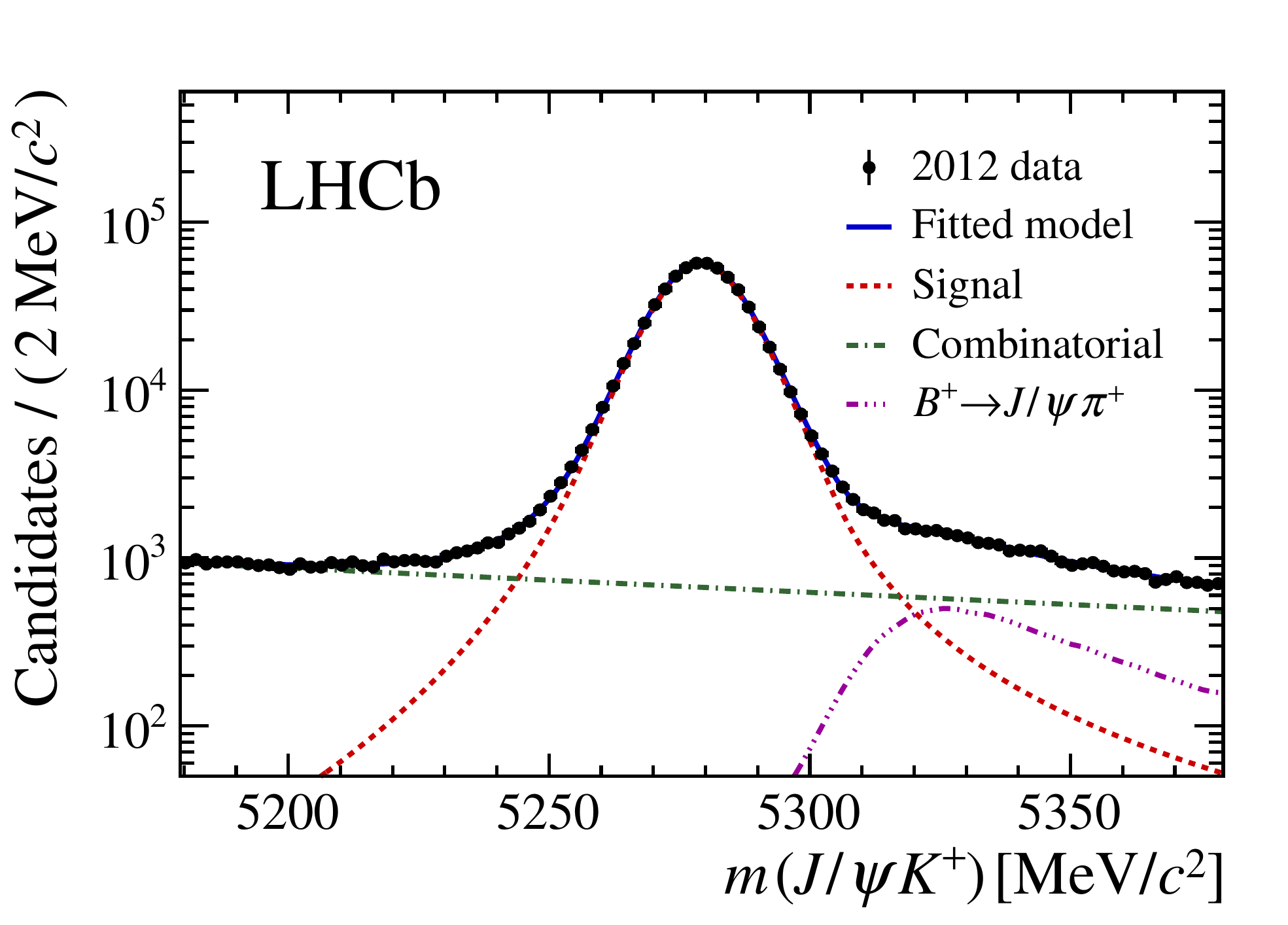}
    \end{subfigure}          
    \begin{subfigure}[t]{0.40\linewidth} 
    \includegraphics[width = \linewidth, trim = 0. 0. 0. 0.]{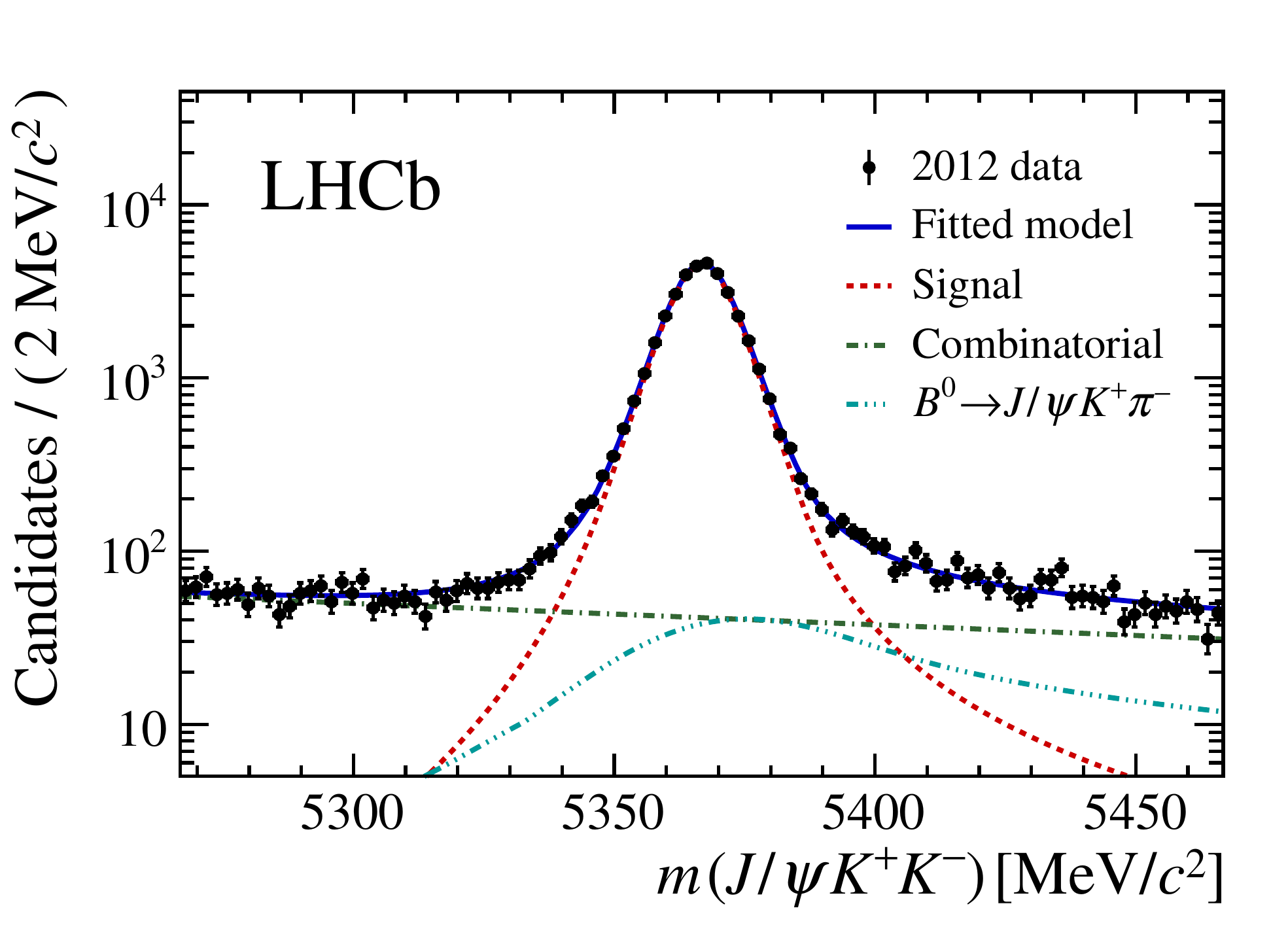}
    \end{subfigure}\\
    \begin{subfigure}[t]{0.40\linewidth} 
    \includegraphics[width = \linewidth, trim = 0. 0. 0. 0.]{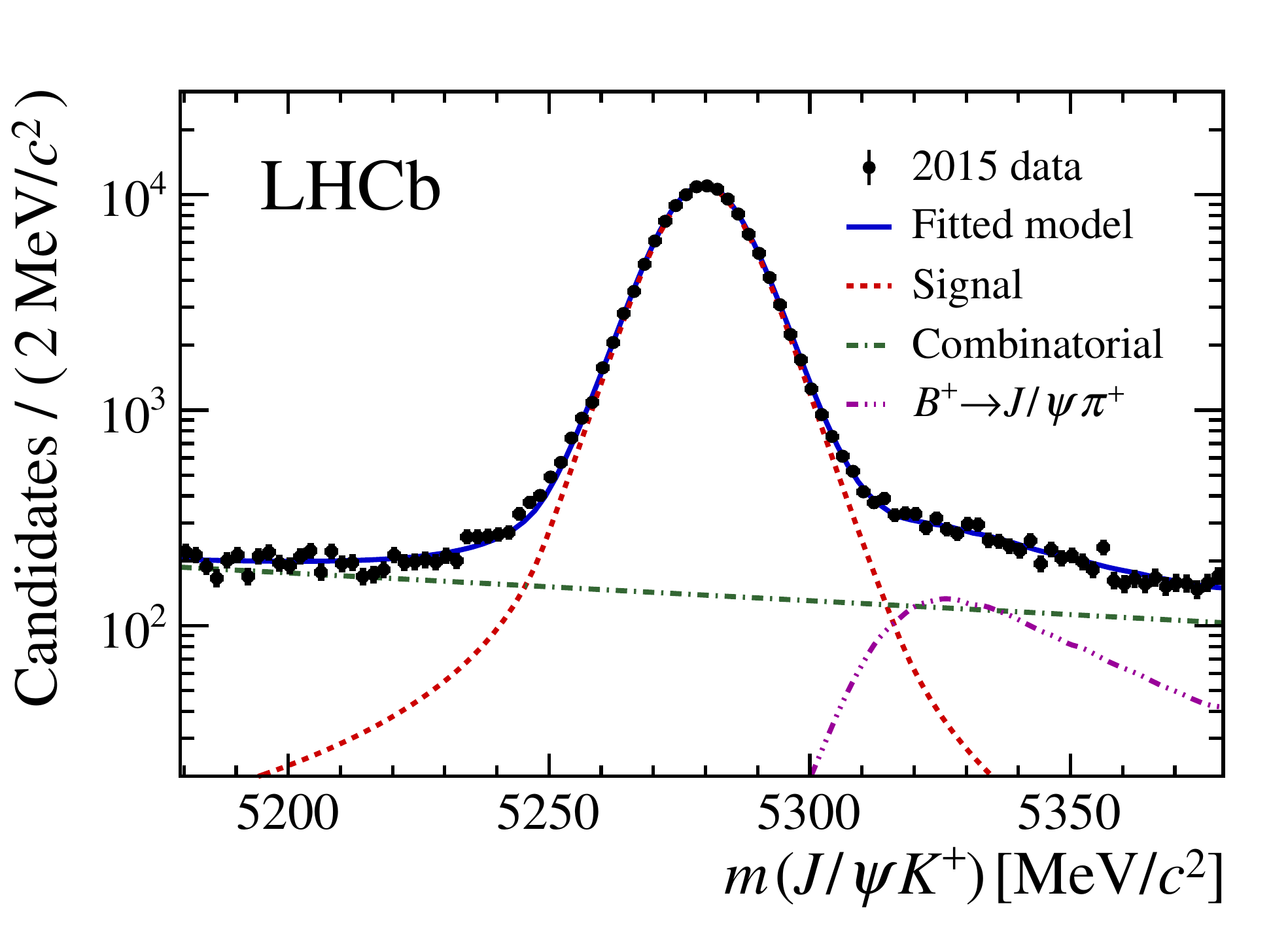}
    \end{subfigure}          
    \begin{subfigure}[t]{0.40\linewidth} 
    \includegraphics[width = \linewidth, trim = 0. 0. 0. 0.]{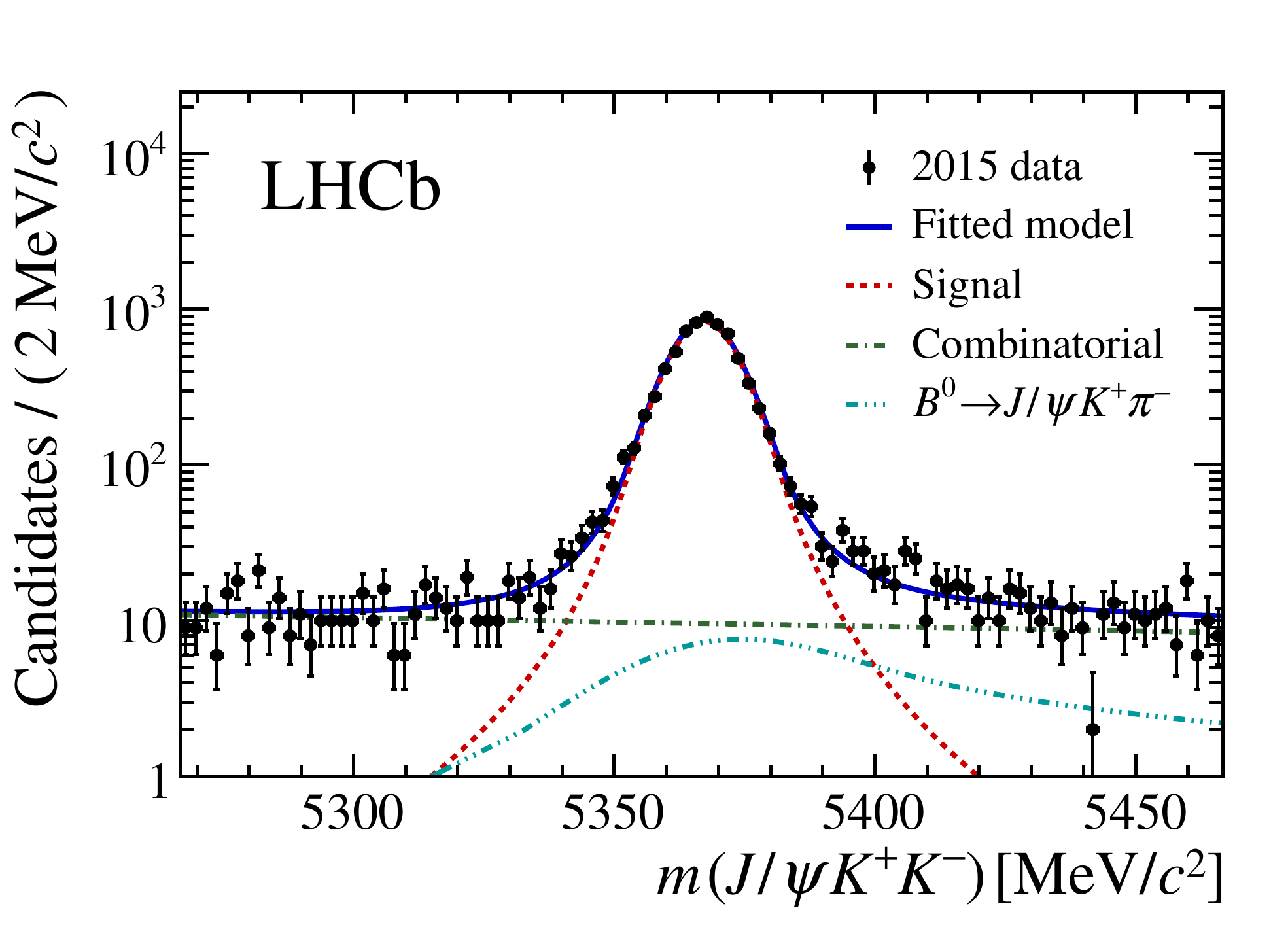}
    \end{subfigure}\\
    \begin{subfigure}[t]{0.40\linewidth} 
    \includegraphics[width = \linewidth, trim = 0. 0. 0. 0.]{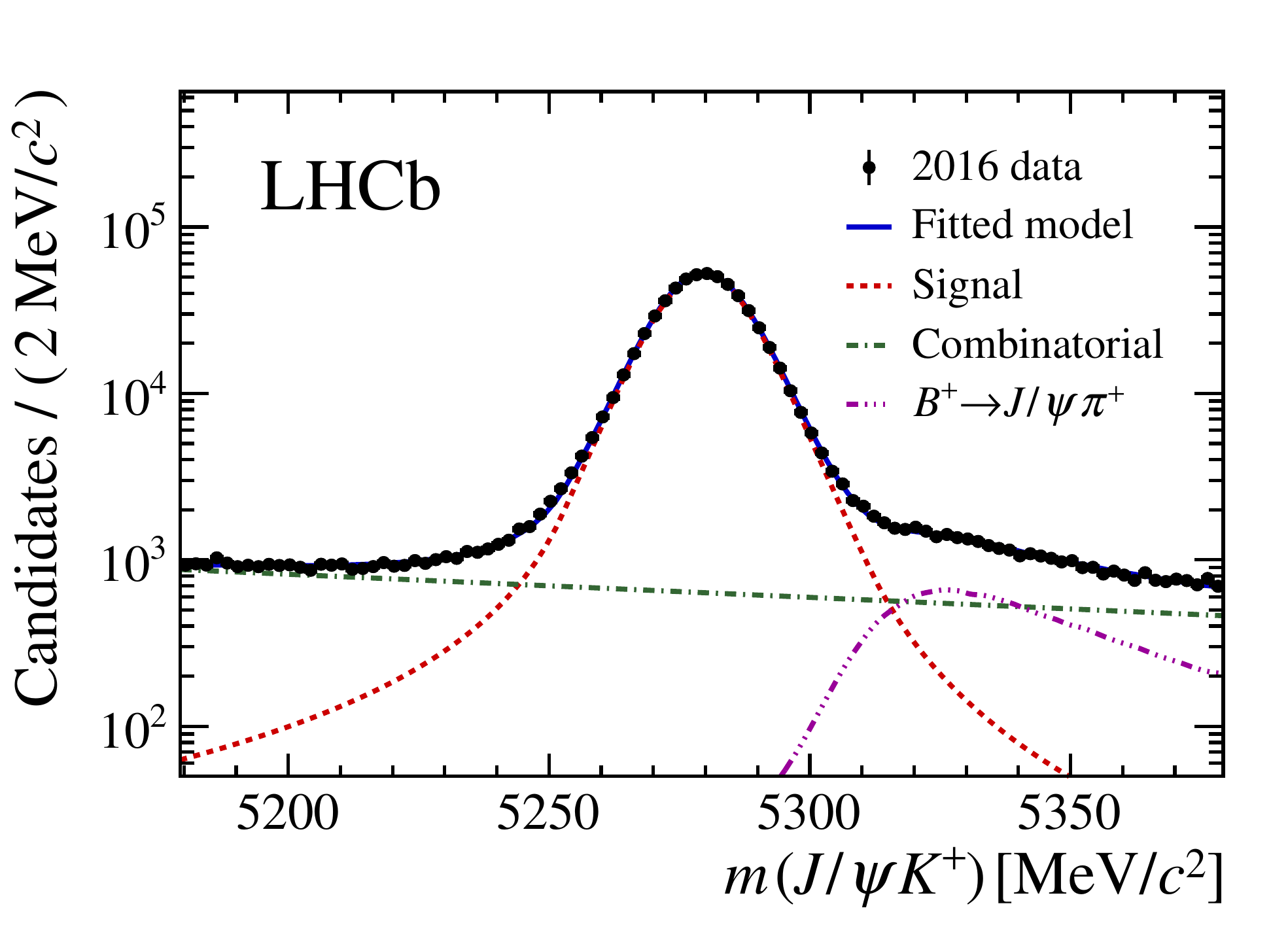}
    \end{subfigure}          
    \begin{subfigure}[t]{0.40\linewidth} 
    \includegraphics[width = \linewidth, trim = 0. 0. 0. 0.]{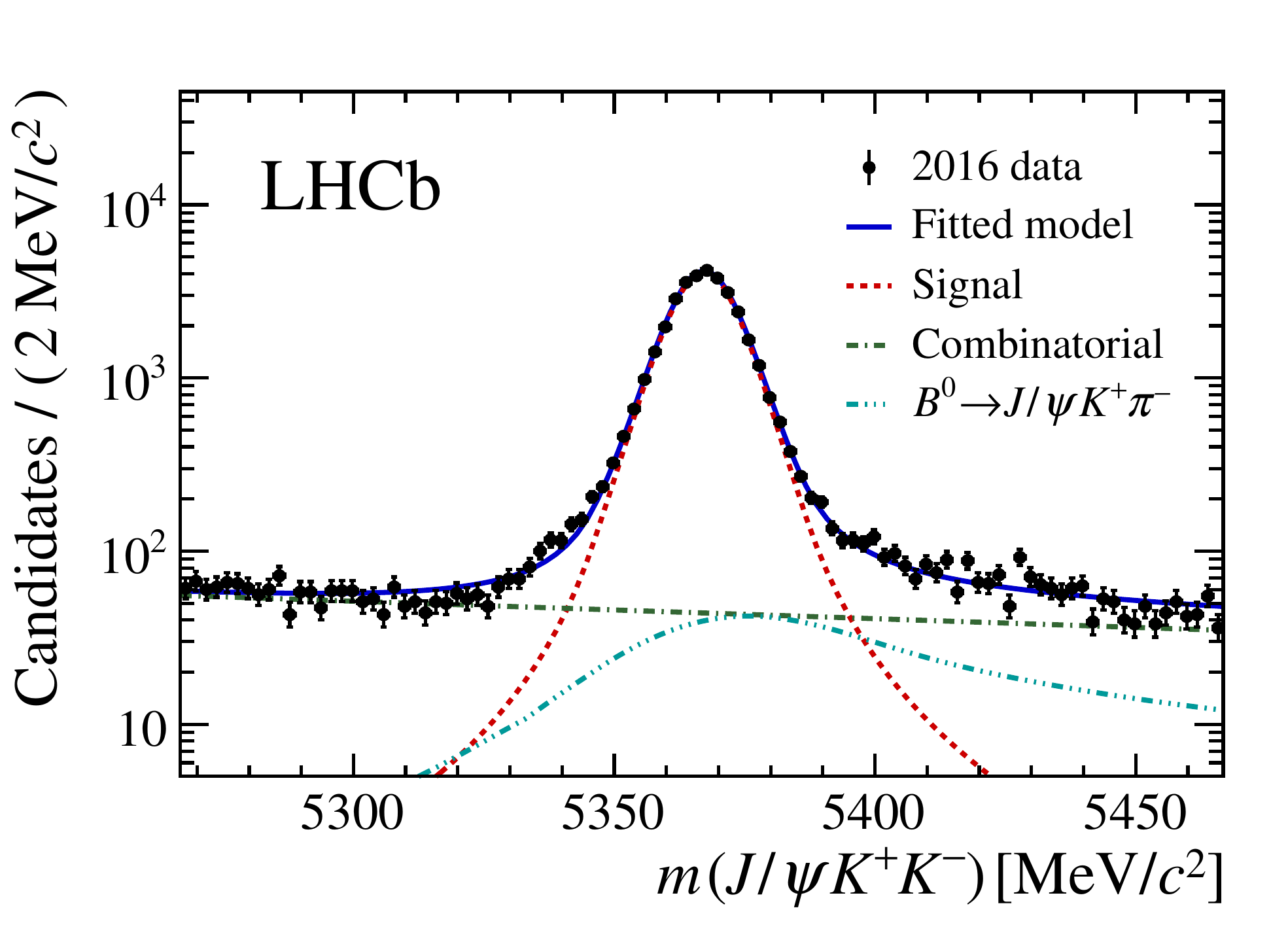}
    \end{subfigure}
    \caption{\footnotesize The \B-meson mass distributions of (left column) \BuJpsiK  and (right column) \BsJpsiPhi candidates in LHCb data collected in 2011, 2012, 2015, and 2016, shown from top to bottom in that order. 
    The  result of the fit is drawn with a blue solid line. The model components are denoted with the dashed lines: signal in red,
    combinatorial background in green, misidentified \bujpsipi in magenta and the misidentified inclusive \bdjpsikpi contribution in light blue.}
    \label{fig:mass2011}
    \end{center} 
\end{figure}

\newpage
\subsection{\boldmath Plotted ratios in bins of \Bptot and \By}
\begin{figure}[hb!]
    \begin{center} 
    \begin{subfigure}[t]{0.49\linewidth} 
    \includegraphics[width = \linewidth, trim = 0. 0. 0. 0.]{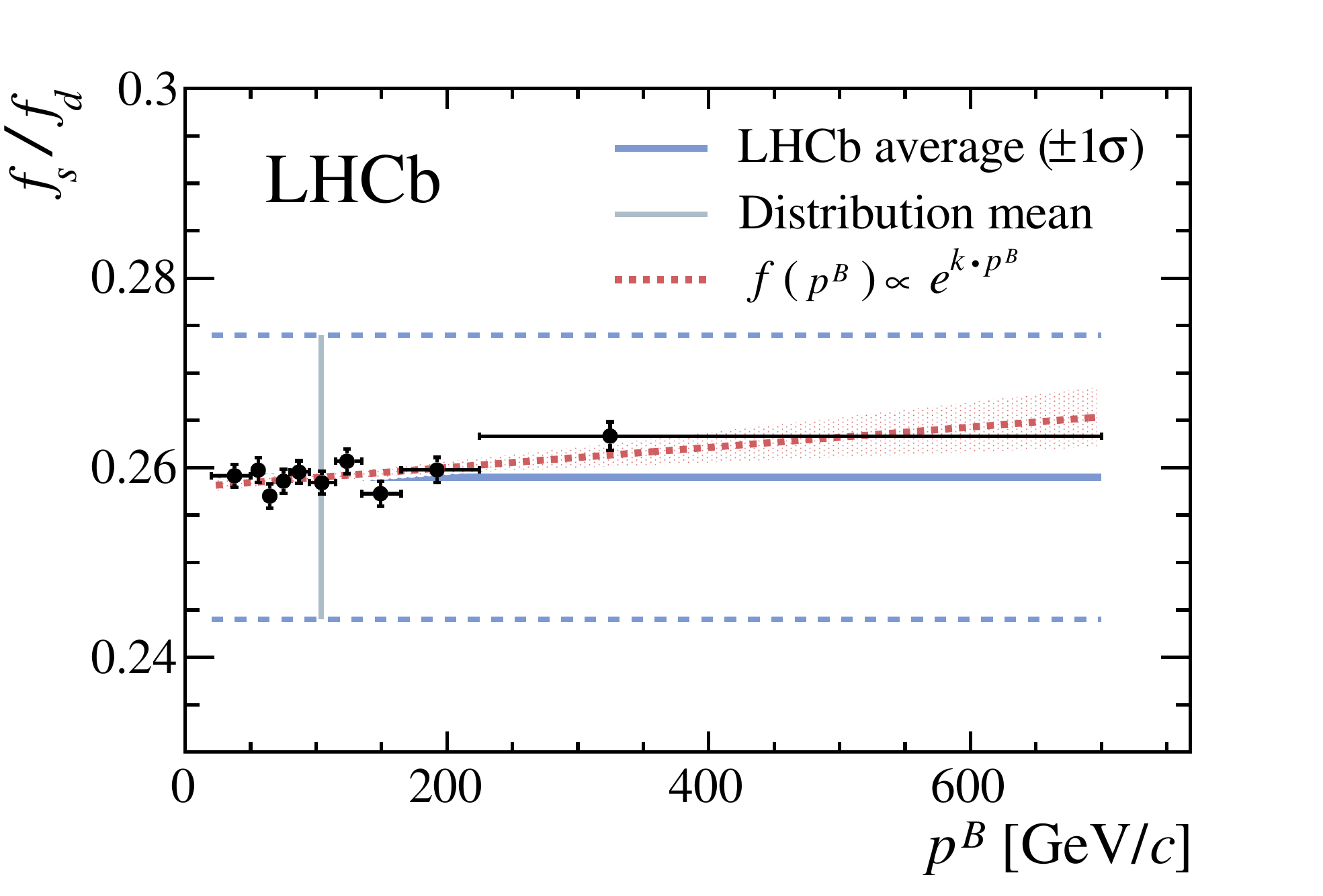}
    \end{subfigure}          
    \begin{subfigure}[t]{0.49\linewidth} 
    \includegraphics[width = \linewidth, trim = 0. 0. 0. 0.]{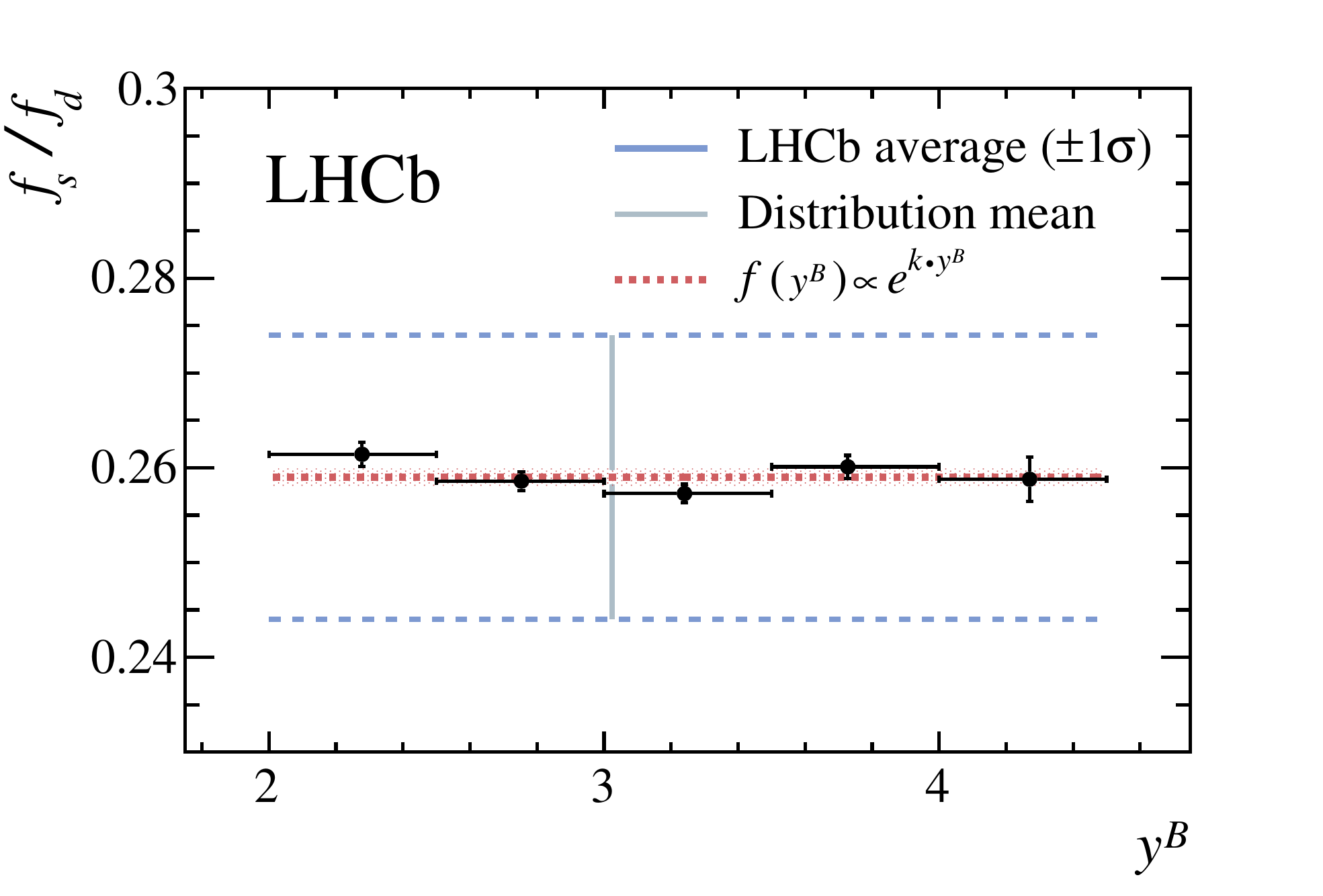}
    \end{subfigure}          
    
    \caption{\small 
    Efficiency-corrected \bsjpsiphi and \bujpsik yield ratios (\Ry) in bins of
    \B-meson momentum \Bptot (left) and rapidity \By (right). 
    The ratios are scaled 
    to match the measured \fsfd value (horizontal blue lines, the $\pm1\sigma$ interval is indicated by the dashed blue lines) at the positions indicated by the vertical gray lines.
    The red dashed line denotes the result of the exponential fit used to estimate the statistical significance of the variation (see text), and the red band denotes the $\protect 68\%$ confidence region. }
    \label{fig:smooth}
    \end{center} 
\end{figure}

\subsection{\boldmath Numerical and plotted ratios in bins of \Bpt for \Bpz subregions}
\begin{table}[hb!]
    \footnotesize
        \caption{\small The efficiency-corrected yield ratio ($\mathcal{R}$) in bins of \B-meson transverse momentum in 
        (a) low, (b) medium, and (c) high \B-meson longitudinal-momentum regions.
        Uncertainties include both statistical and systematic sources.} 
    \begin{center}

\begin{subtable}[t]{\textwidth}
    \begin{tabular}{lc} 
        \multicolumn{2}{c}{a) \footnotesize $ 0 \leq\Bpz < 75 \gevc$} \\
        \\ 
        \toprule
        Range [\gevc] &  $\mathcal{R}$\\
        \midrule 
        $0.5 < \Bpt < 4 $   & $0.124 \pm 0.002$ \\
        $4 <   \Bpt <6$     & $0.127 \pm 0.003$ \\
        $6 <   \Bpt <8$     & $0.129 \pm 0.003$ \\
        $8 <   \Bpt <11$    & $0.125 \pm 0.004$ \\
        $11 <   \Bpt <40$   & $0.119 \pm 0.006$ \\ 
        \bottomrule
    \end{tabular}    
        %\end{subtable}\hspace{0 cm}
        %\begin{subtable}[t]{0.35\textwidth}
        %\caption{Results as a function of the transverse momentum in low longitudinal momentum region: $ 75 \leq\Bpz < 125 \gevc$.}
    \begin{tabular}{lc} 
        \multicolumn{2}{c}{b) \footnotesize $ 75 \leq\Bpz < 125 \gevc$} \\
        \\ 
        \toprule
        Range [\gevc] &  $\mathcal{R}$\\
        \midrule 
        %[250, 4000., 6000., 8000., 11000., 40000.]\me        
        $0.5 < \Bpt < 4 $  & $0.128 \pm 0.003$ \\
        $4 <   \Bpt <6$     & $0.132 \pm 0.003$ \\
        $6 <   \Bpt <8$     & $0.128 \pm 0.003$ \\
        $8 <   \Bpt <11$    & $0.129 \pm 0.003$ \\
        $11 <   \Bpt <40$   & $0.119 \pm 0.003$ \\
        \bottomrule
    \end{tabular}    
    %\end{subtable}\hspace{0 cm}
    %\begin{subtable}[t]{0.35\textwidth}
    \begin{tabular}{lc} 
        \multicolumn{2}{c}{c) \footnotesize $ 125 \leq\Bpz < 700 \gevc$} \\
        \\ 
        \toprule
        Range [\gevc] &  $\mathcal{R}$\\
        \midrule 
        %[250, 4000., 6000., 8000., 11000., 40000.]\me        
        $0.5 < \Bpt < 4 $  & $0.131 \pm 0.004$ \\
        $4 <   \Bpt <6$     & $0.128 \pm 0.004$ \\
        $6 <   \Bpt <8$     & $0.127 \pm 0.003$ \\
        $8 <   \Bpt <11$    & $0.123 \pm 0.003$ \\
        $11 <   \Bpt <40$   & $0.121 \pm 0.002$ \\
        \bottomrule
    \end{tabular}    
    \end{subtable}
\end{center}
\end{table}

\begin{figure}[tb!]
    \begin{center} 
    
    \begin{subfigure}[t]{0.49\linewidth} 
    \includegraphics[width = \linewidth, trim = 0. 0. 0. 0.]{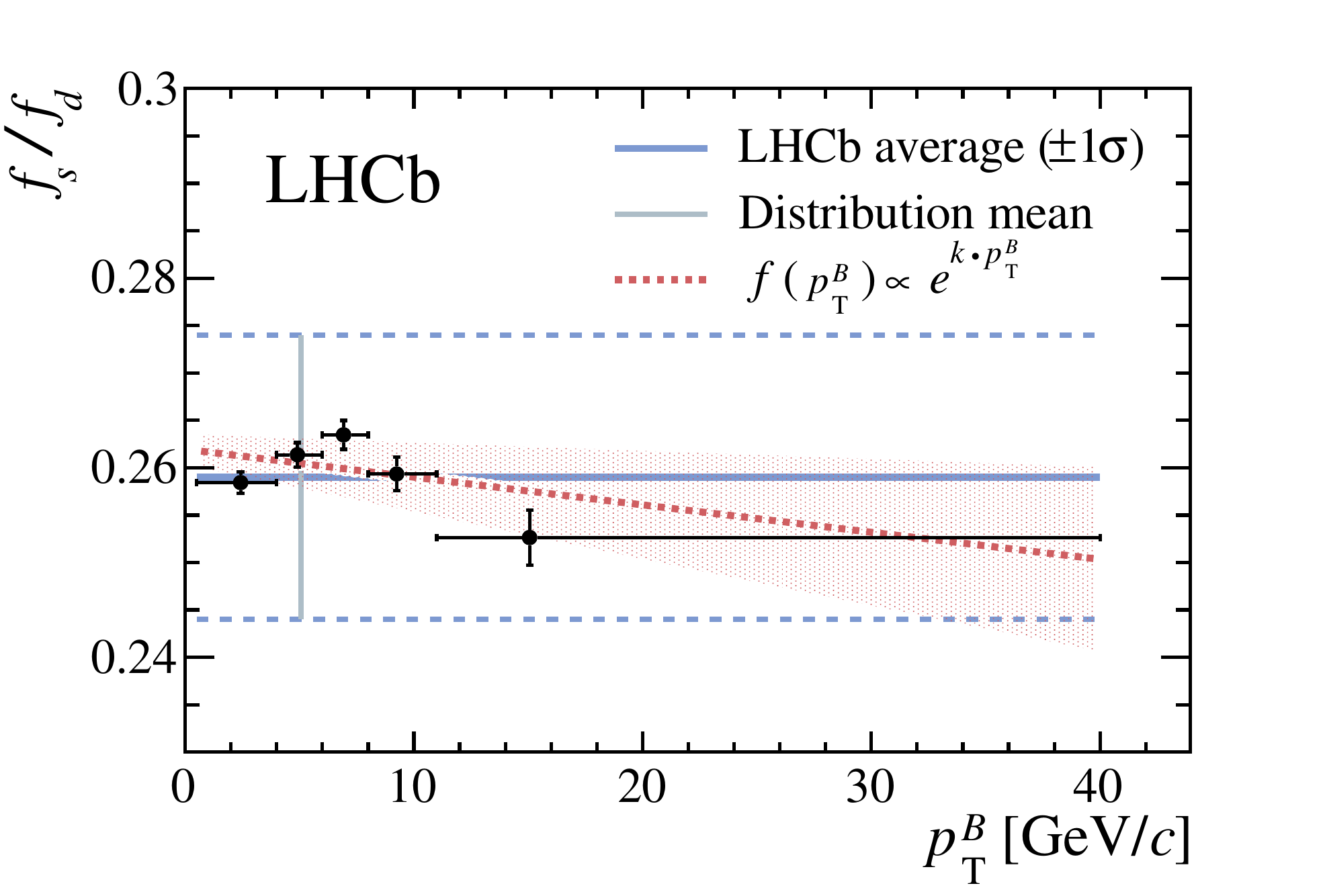}
    \end{subfigure}          
    
    \begin{subfigure}[t]{0.49\linewidth} 
    \includegraphics[width = \linewidth, trim = 0. 0. 0. 0.]{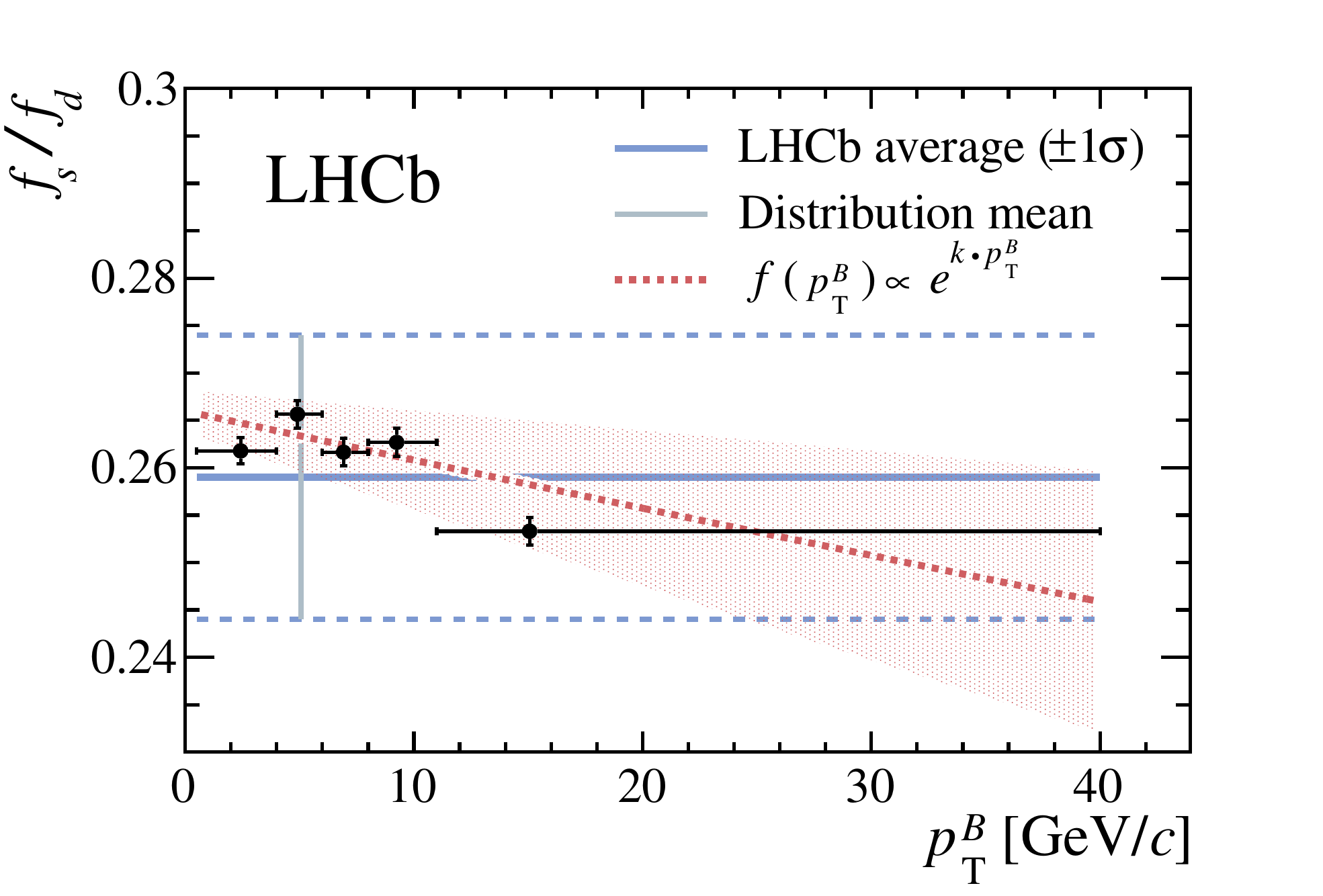}
    \end{subfigure}          
    
    \begin{subfigure}[t]{0.49\linewidth} 
    \includegraphics[width = \linewidth, trim = 0. 0. 0. 0.]{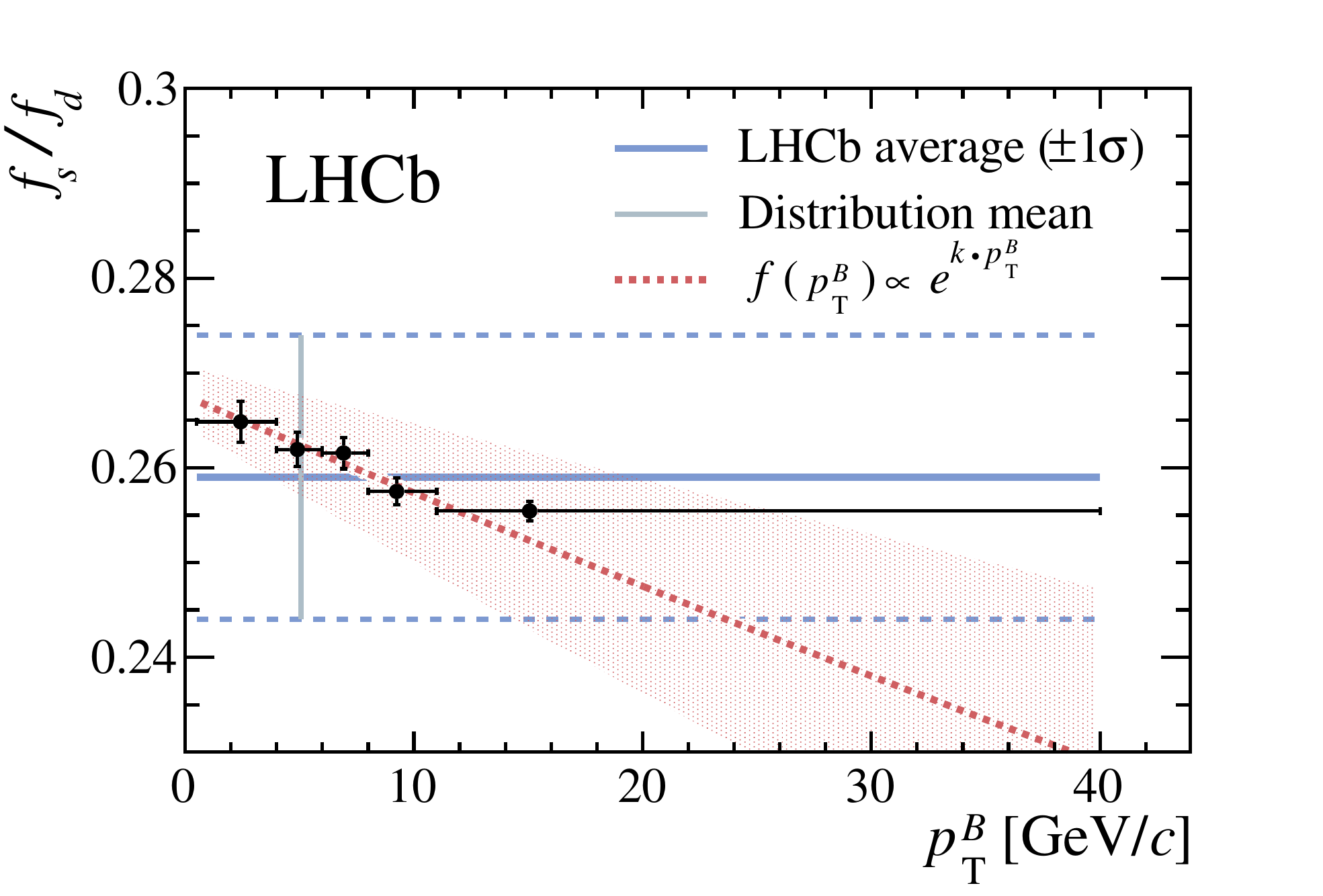}
    \end{subfigure}

    \caption{\small 
    Efficiency-corrected \bsjpsiphi and \bujpsik yield ratios (\Ry) in bins of \B-meson transverse momentum \Bpt,
    shown for the \B-meson longitudinal-momentum ranges:  
    (top) low range ([0,75)\gevc), 
    (middle) medium range ([75,125)\gevc), and 
    (bottom) high range ([125,700]\gevc). 
    The ratios are scaled 
    to match the measured \fsfd value (horizontal blue lines, the $\pm1\sigma$ interval is indicated by the dashed blue lines) at the positions indicated by the vertical gray lines.
    The red dashed line denotes the result of the exponential fit used to estimate the variation (see text), and the red band denotes the $\protect 68\%$ confidence region.}
    \label{fig:variations}
    \end{center} 
\end{figure}

\clearpage
\subsection{\boldmath Numerical ratios in bins of \Bptot, \Bpz, \Bpt, \Beta, and \By}

\begin{table}[b!]
    \footnotesize
    \caption{\small The measured efficiency-corrected yield ratio ($\mathcal{R}$) in bins of the kinematic variables. Uncertainties include both statistical and systematic sources.} 
    \begin{center}
    \begin{subtable}[t]{0.35\textwidth}
    \caption{Results as a function of the total \B-meson momentum. }
    \begin{tabular}{lc} 
    \toprule
   Range [\gevc] &  $\mathcal{R}$\\
\midrule 
% \item \Bptot    = [0.0, 50000, 60000, 70000., 80000., 95000., 115000., 135000., 165000., 225000., 700000.] \mevc
  $20  <\Bptot <50$         & $0.127 \pm 0.002$ \\
  $50 < \Bptot <60$         & $0.127 \pm 0.003$ \\
  $60 < \Bptot< 70$         & $0.125 \pm 0.003$ \\
  $70 < \Bptot <80 $        & $0.126 \pm 0.003$ \\
  $80 < \Bptot < 95$        & $0.127 \pm 0.002$ \\
  $95  <\Bptot < 110$       & $0.126 \pm 0.002$ \\
  $110 < \Bptot <135$       & $0.128 \pm 0.003$ \\
  $135 < \Bptot <165$       & $0.125 \pm 0.003$ \\
  $165  <\Bptot <225$       & $0.127 \pm 0.003$ \\
  $225  <\Bptot <700$       & $0.131 \pm 0.003$ \\
\midrule 
\end{tabular}    \end{subtable}\hspace{1 cm}
\begin{subtable}[t]{0.35\textwidth}
    \caption{Results as a function of the \B-meson longitudinal momentum.}
    \begin{tabular}{lc} 
    \toprule 
   Range [\gevc] &  $\mathcal{R}$\\
\midrule 

% \item \Bpz      = [0.0, 50000., 60000., 70000., 80000., 95000., 110000., 135000., 165000., 225000., 700000.] \mevc
$20  < \Bpz <  50$      & $0.126 \pm 0.002$ \\
$50  < \Bpz <  60$      & $0.127 \pm 0.003$ \\
$60  < \Bpz <  70$      & $0.125 \pm 0.003$ \\
$70  < \Bpz <  80$      & $0.127 \pm 0.003$ \\
$80  < \Bpz <  95$      & $0.127 \pm 0.002$ \\
$95  < \Bpz <  110$     & $0.127 \pm 0.003$ \\
$110  < \Bpz <  135$    & $0.127 \pm 0.002$ \\
$135  < \Bpz <  165$    & $0.125 \pm 0.003$ \\
$165  < \Bpz <  225$    & $0.127 \pm 0.003$ \\
$225  < \Bpz <  700$    & $0.130 \pm 0.003$ \\
        \bottomrule                                                               
    \label{tab:results_fsfds_p}
    \end{tabular}
    \end{subtable}\\[10pt]

\begin{subtable}[t]{0.35\textwidth}
\caption{Results as a function of the \B-meson transverse momentum.}
    \begin{tabular}{lc} 
        \toprule
   Range [\gevc] &  $\mathcal{R}$\\
\midrule 
% \item \Bpt      = [0.0, 2000, 3000, 4000, 5000, 6000, 7000, 8000, 9000, 10000, 11500, 14000, 40000] \mevc
$0.5 < \Bpt < 2 $     & $0.125 \pm 0.003$ \\
$2 <   \Bpt <3$        & $0.127 \pm 0.003$ \\
$3 <   \Bpt <4$        & $0.125 \pm 0.003$ \\
$4 <   \Bpt <5$        & $0.128 \pm 0.003$ \\
$5 <   \Bpt <6$        & $0.128 \pm 0.003$ \\
$6 <   \Bpt <7$        & $0.127 \pm 0.003$ \\
$7 <   \Bpt <8$        & $0.127 \pm 0.003$ \\
$8 <   \Bpt <9$        & $0.126 \pm 0.003$ \\
$9 <   \Bpt <10$       & $0.125 \pm 0.003$ \\
$10 <   \Bpt < 11.5$   & $0.125 \pm 0.003$ \\
$11.5 <  \Bpt <14$     & $0.118 \pm 0.003$ \\
$14 <   \Bpt < 40$     & $0.120 \pm 0.002$ \\
        \bottomrule
\end{tabular}    
\end{subtable}\hspace{1 cm}
    \begin{subtable}[t]{0.35\textwidth}
    \caption{Results as a function of the \B-meson pseudorapidity.}
             \begin{tabular}{lc} 
    \toprule
    Range       & $\mathcal{R}$\\
\midrule
    $2.0 < \Beta < 2.5$      & $0.127 \pm 0.004$ \\
    $2.5 < \Beta < 2.8$      & $0.131 \pm 0.003$ \\
    $2.8 <\Beta < 3.0$       & $0.129 \pm 0.003$ \\
    $3.0 <\Beta <3.2$        & $0.130 \pm 0.002$ \\
    $3.2 < \Beta <3.4$       & $0.126 \pm 0.002$ \\
    $3.4 <\Beta <3.6$        & $0.125 \pm 0.002$ \\
    $3.6 <\Beta <3.8$        & $0.127 \pm 0.002$ \\
    $3.8 <\Beta< 4.0$        & $0.128 \pm 0.003$ \\
    $4.0 < \Beta < 4.3$      & $0.129 \pm 0.003$ \\
    $4.3 < \Beta < 6.4$      & $0.130 \pm 0.002$ \\ 
    \bottomrule
    \label{tab:results_fsfds}
    \end{tabular}
        \end{subtable}\\[10pt]
        
    \begin{subtable}[t]{0.35\textwidth}
    \caption{Results as a function of the \B-meson rapidity.}
    \begin{tabular}{lc}
    \toprule
    Range &  $\mathcal{R}$\\
\midrule 
$2.0 <\By < 2.5 $  & $0.130 \pm 0.003$ \\
$2.5<\By <  3.0$   & $0.127 \pm 0.002$ \\
$3.0<\By <  3.5$  & $0.126 \pm 0.002$ \\
$3.5<\By <  4.0$  & $0.128 \pm 0.003$ \\
$4.0<\By <  4.5$  & $0.127 \pm 0.005$ \\
        \hline                                                                 
    \label{tab:results_fsfds_etay}
    \end{tabular}
\end{subtable}\hspace{1cm}\hspace{0.35\textwidth}

    \end{center}
\end{table}

\newpage
\subsection{\boldmath Plotted efficiency-corrected yield ratios as a function of \deltay}

Given the availability of data at different center-of-mass energies, 
results can be compared as a function of the variable
\begin{equation*}
  \deltay = y_{\rm{beam}} - y^{B}
\end{equation*}
where $y_{\rm{beam}}$ is the rapidity of the incoming proton beam and \By the \B-meson rapidity. 
This variable is typically defined in association with the transport of the baryon number
from the initial to the final state in $pp \to N X$ reactions, where $N$ is a generic baryon and $X$ can be any accompanying process. 
However, it can be useful also to understand the hadronization process for mesons. 

\begin{figure}[b!]
\begin{center}    \includegraphics[width=0.7\linewidth, trim = 0. 0. 0. 0.]{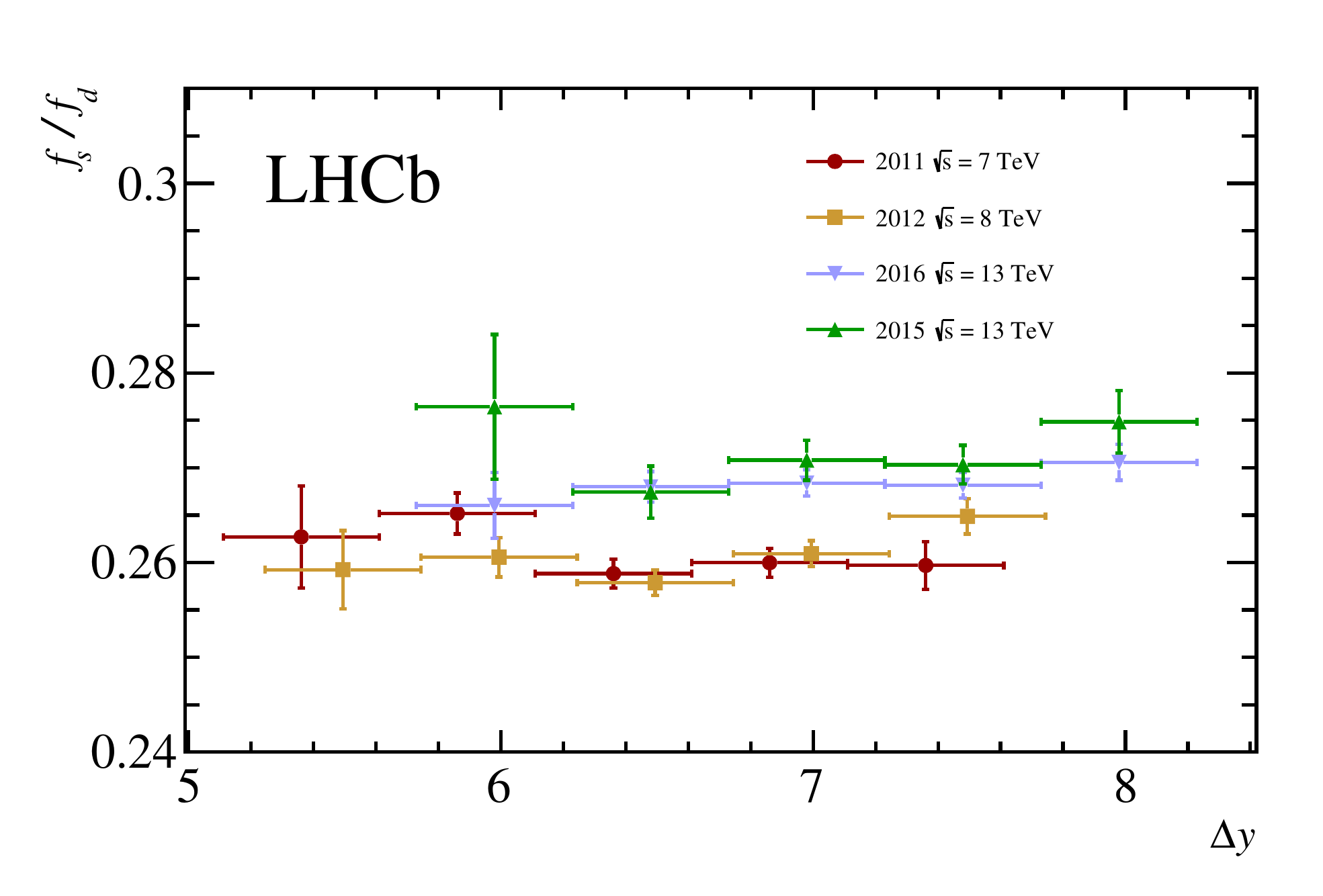} 
 \end{center}
   \caption{\footnotesize 
    Efficiency-corrected \bsjpsiphi and \bujpsik yield ratios (\Ry) in bins of $\deltay$ for different samples. 
    The ratios are scaled to match the measured \fsfd value.}
    \label{fig:fsfd_in_deltay}
\end{figure}

The results of the efficiency-corrected yield ratios as a function of \deltay are shown in Fig.~\ref{fig:fsfd_in_deltay}
as obtained by shifting those as a function of \By by the corresponding $y_{\rm beam}$. 
This variable is useful for comparison with ATLAS and CMS experiments. 
As an example, LHCb data at $\sqrt{s} = 13\tev$ (or $y_{\rm{beam}} = 10.2$) and rapidity $y\simeq 2$
could be compared with ATLAS/CMS data at $\sqrt{s} = 7 \tev$ ($y_{\rm{beam}} = 9.6$) and rapidity $y\simeq 1$; a region otherwise unavailable to LHCb at $\sqrt{s} = 7\tev$. 

\newpage
\subsection{\boldmath Numerical and plotted ratios in bins of \Bpt for different \pp collision energies}

\begin{table}[h]
    \footnotesize
    \caption{\small Efficiency-corrected \bsjpsiphi and \bujpsik yield ratios in bins of \B-meson transverse momentum \Bpt, separately for the three \pp collision energies.} 
    \begin{center}
    \begin{subtable}[t]{0.35\textwidth}
    \caption{Results at $\sqrt{s} = 7~\tev$}
    \begin{tabular}{lc} 
    \toprule
   Range [\gevc] &  $\mathcal{R}$\\
\midrule 
$0.5 <\Bpt < 2 $     & $0.119 \pm 0.003$ \\
$2 <\Bpt < 3 $     & $0.127 \pm 0.003$ \\
$3 <\Bpt < 4 $     & $0.120 \pm 0.002$ \\
$4 <\Bpt < 5 $     & $0.122 \pm 0.002$ \\
$5 <\Bpt < 6 $     & $0.122 \pm 0.002$ \\
$6 <\Bpt < 7 $     & $0.128 \pm 0.003$ \\
$7 <\Bpt < 8 $     & $0.129 \pm 0.003$ \\
$8 <\Bpt < 9 $     & $0.129 \pm 0.003$ \\
$9 <\Bpt < 10 $     & $0.115 \pm 0.003$ \\
$10 <\Bpt < 12 $     & $0.116 \pm 0.003$ \\
$11 <\Bpt < 14 $     & $0.118 \pm 0.003$ \\
$14 <\Bpt < 40 $     & $0.117 \pm 0.003$ \\

\midrule 
\end{tabular}    \end{subtable}\hspace{1 cm}
\begin{subtable}[t]{0.35\textwidth}
    \caption{Results at $\sqrt{s} = 8~\tev$}
    \begin{tabular}{lc} 
    \toprule 
   Range [\gevc] &  $\mathcal{R}$\\
\midrule 
$0.5 <\Bpt < 2 $     & $0.121 \pm 0.002$ \\
$2 <\Bpt < 3 $     & $0.121 \pm 0.002$ \\
$3 <\Bpt < 4 $     & $0.120 \pm 0.002$ \\
$4 <\Bpt < 5 $     & $0.127 \pm 0.002$ \\
$5 <\Bpt < 6 $     & $0.125 \pm 0.002$ \\
$6 <\Bpt < 7 $     & $0.125 \pm 0.002$ \\
$7 <\Bpt < 8 $     & $0.121 \pm 0.002$ \\
$8 <\Bpt < 9 $     & $0.122 \pm 0.002$ \\
$9 <\Bpt < 10 $     & $0.125 \pm 0.002$ \\
$10 <\Bpt < 12 $     & $0.125 \pm 0.002$ \\
$11 <\Bpt < 14 $     & $0.119 \pm 0.002$ \\
$14 <\Bpt < 40 $     & $0.119 \pm 0.002$ \\
        \bottomrule                                                               
    \label{tab:results_fsfds_pt}
    \end{tabular}
    \end{subtable}\\[10pt]

\begin{subtable}[t]{0.35\textwidth}
\caption{Results at $\sqrt{s} = 13~\tev$.}
    \begin{tabular}{lc} 
        \toprule
   Range [\gevc] &  $\mathcal{R}$\\
\midrule 
$0.5 <\Bpt < 2 $     & $0.133 \pm 0.002$ \\
$2 <\Bpt < 3 $     & $0.132 \pm 0.002$ \\
$3 <\Bpt < 4 $     & $0.134 \pm 0.002$ \\
$4 <\Bpt < 5 $     & $0.132 \pm 0.002$ \\
$5 <\Bpt < 6 $     & $0.134 \pm 0.002$ \\
$6 <\Bpt < 7 $     & $0.129 \pm 0.002$ \\
$7 <\Bpt < 8 $     & $0.131 \pm 0.002$ \\
$8 <\Bpt < 9 $     & $0.129 \pm 0.002$ \\
$9 <\Bpt < 10 $     & $0.128 \pm 0.002$ \\
$10 <\Bpt < 12 $     & $0.128 \pm 0.002$ \\
$11 <\Bpt < 14 $     & $0.118 \pm 0.002$ \\
$14 <\Bpt < 40 $     & $0.121 \pm 0.002$ \\
        \bottomrule
\end{tabular}    
\end{subtable}\hspace{1 cm}
\hspace{0.35\textwidth}
    \end{center}
\end{table}

% \begin{figure}[p!]
% \begin{center}    \includegraphics[width=0.7\linewidth, trim = 0. 0. 0. 0.]{Fig8_Sup.pdf} 
%  \end{center}
%    \caption{\footnotesize 
%     Efficiency-corrected \bsjpsiphi and \bujpsik yield ratios (\Ry) in bins of \B-meson transverse momentum \Bpt. 
%     The ratios are scaled to match the measured \fsfd value (horizontal blue lines, the $\pm1\sigma$ interval is 
%     indicated by the dashed blue lines) at the positions indicated by the vertical gray lines.
%     The same scale factor is used for all the samples.
%     }
%     \label{fig:fsfd_in_pt_energies}
% \end{figure}

The statistical significance of the \fsfu dependence on \Bpt at each \pp collision energy is estimated by fitting the efficiency-corrected 
\bsjpsiphi and \bujpsik yield ratios (\Ry, see \tabref{tab:results_fsfds_pt} and Fig. 3b of the main text) distributions with a 
function $A_{\Bpt}\cdot \exp(k_{\Bpt}\cdot \Bpt)$ under two hypotheses: one where 
no variation is allowed and the slope parameter, $k_{\Bpt}$, is fixed to zero and one with $k_{\Bpt}$ left free. 
The $\chisq$ difference between the two cases is used as a test statistic and its $p$-value is determined from the 
$\chisq$ distribution with one degree of freedom~\cite{Wilks:1938dza}.
The results are 
$k_{\Bpt}= -(1.24\pm0.73)\times10^{-3} \gev^{-1}c$ ($2.1\,\sigma$),
$k_{\Bpt}= -(0.59\pm0.39)\times10^{-3} \gev^{-1}c$ ($1.5\,\sigma$) and
$k_{\Bpt}= -(4.40\pm0.67)\times10^{-3} \gev^{-1}c$ ($8.7\,\sigma$)
for $7\tev$, $8\tev$ and $13\tev$ samples, respectively. The two-sided significances are given in the brackets.

%% file: LHCb_Authorship_01-Oct-2019.tex
% LHCb collaboration author list
% Data extracted on October 11th, 2019 at 5:29pm for reference date 01-Oct-2019
\centerline
{\large\bf LHCb collaboration}
\begin
{flushleft}
\small
R.~Aaij$^{31}$,
C.~Abell{\'a}n~Beteta$^{49}$,
T.~Ackernley$^{59}$,
B.~Adeva$^{45}$,
M.~Adinolfi$^{53}$,
H.~Afsharnia$^{9}$,
C.A.~Aidala$^{80}$,
S.~Aiola$^{25}$,
Z.~Ajaltouni$^{9}$,
S.~Akar$^{66}$
% ...

{\footnotesize \it

% ...
$ ^{9}$Universit{\'e} Clermont Auvergne, CNRS/IN2P3, LPC, Clermont-Ferrand, France\\
% ...
$ ^{25}$INFN Sezione di Milano, Milano, Italy\\
% ...
$ ^{31}$Nikhef National Institute for Subatomic Physics, Amsterdam, Netherlands\\
% ...
$ ^{45}$Instituto Galego de F{\'\i}sica de Altas Enerx{\'\i}as (IGFAE), Universidade de Santiago de Compostela, Santiago de Compostela, Spain\\
% ...
$ ^{49}$Physik-Institut, Universit{\"a}t Z{\"u}rich, Z{\"u}rich, Switzerland\\
% ...
$ ^{53}$H.H. Wills Physics Laboratory, University of Bristol, Bristol, United Kingdom\\
% ...
$ ^{59}$Oliver Lodge Laboratory, University of Liverpool, Liverpool, United Kingdom\\
% ...
$ ^{66}$Los Alamos National Laboratory (LANL), Los Alamos, United States\\
$ ^{67}$Syracuse University, Syracuse, NY, United States\\
% ...
$ ^{80}$University of Michigan, Ann Arbor, United States, associated to $^{67}$\\
\bigskip
% ...
}
\end{flushleft}